\documentclass[nopubldata]{lpor2014}

\usepackage{siunitx}
\usepackage{color}
\usepackage{overpic}
\usepackage{tabularx}
\usepackage[ampersand]{easylist}

\newcommand{\inj}{_\mathrm{inj}}

\usepackage[overlay,absolute]{textpos}
\newcommand\PlaceText[3]{
	\begin{textblock*}{10in}(#1,#2)
		#3
	\end{textblock*}
}

\begin{document}
	\abstract{Quantum communications is the art of exchanging and manipulating information beyond the capabilities of our conventional technologies using the laws of quantum mechanics. With applications ranging from quantum computing to cryptographic systems with information-theoretic security, there is strong incentive to introduce quantum communications into many areas of our society. However, an important challenge is to develop viable technologies meeting the stringent requirements of low noise and high coherence for quantum state encoding, of high bit rate and low power for the integration with classical communication networks and of scalable and low-cost production for a practical wide-deployment. This tutorial presents recent advances in laser modulation technologies that have enabled the development of efficient and versatile light sources for quantum communications, with a particular focus on quantum key distribution (QKD). Such approaches have been successfully used to demonstrate several QKD protocols with state-of-the-art performance. The applications and experimental results are reviewed and interpreted in the light of a complete theoretical background, allowing the reader to model and simulate such sources.}
	
	\title{Advanced Laser Technology for Quantum Communications (\emph{Tutorial Review})}
	
	\titlerunning{Advanced laser technology for quantum communications}
	
	\author{T.~K.~Para\"iso$^{1,*}$, R.~I.~Woodward$^1$, D.~G.~Marangon$^1$, V. Lovic$^{1,2}$, Z.~L.~Yuan$^1$ and A.~J.~Shields$^1$}
	
	\authorrunning{T.K. Para\"iso et al.}
	
	\institute{%
		$^1$ Toshiba Europe Ltd, Cambridge, UK\\
		$^2$ QOLS, Blackett Laboratory, Imperial College London, UK\\
	}

	\mail{\email{taofiq.paraiso@crl.toshiba.co.uk}}

	
	\maketitle
	
	\PlaceText{12mm}{8mm}{Adv. Quantum Technol. \textbf{4}, 2100062 (2021); https://doi.org/10.1002/qute.202100062}
	
	\section{Introduction}
	\label{sec:intro}
	
	Quantum communications, i.e. the encoding and transfer of quantum states between distant parties, is set to occupy a central place in the future of information exchange and processing. In particular, owing to the threat of quantum computers against conventional public-key cryptography algorithms, quantum key distribution (QKD) offers means of securely establishing symmetric encryption keys with a security level quantifiable using information theory \cite{Gisin2002}. QKD has been developing at an ever increasing pace over the last two decades, and the recent years have been marked by impressive results establishing the maturity of the technology \cite{FeihuXu.2020}. Notable successes include resilience to attacks on classical hardware \cite{Lydersen2010,Yuan2010,huang_laser-seeding_2019,Koehler-Sidki2018,M.Lucamarini.2015}, high bit rate secure key distribution \cite{Yuan.2018,Islam2017}, long-distance QKD links \cite{Lucamarini.2018,Pittaluga.2020,Chen2021,Boaron.2018,ShouvikGhorai.2019,YichenZhang.2020}, few-node networks \cite{Sasaki.2011,Stucki2011,Dynes2019} and more recently, satellite-borne QKD \cite{Liao.2017} and hybrid space-to-ground networks \cite{YuAoChen.2021}. This high level of maturity has motivated governments, research institutions and industry partners to develop large-scale quantum communication infrastructures and to standardize the technology in order to enable its practical integration with classical fiber communication networks \cite{Tanizawa2017,Mao2018,ETSI2019a}. Exhaustive reviews of the progress in QKD can be found in Refs. \cite{FeihuXu.2020,Pirandola2019}
	
	The perspective of large-scale deployment of quantum communication technologies poses an immediate practicality challenge: \emph{how to encode quantum information in a versatile way while preserving the constraints of low power budget, small size and high scalability required for a realistic integration in our conventional communication infrastructure} \cite{Diamanti.2016}?
	We note that this question is relevant to applications beyond quantum cryptography: while quantum communications is often presented as antagonist to quantum computing, the two fields become complementary in the perspective of a quantum internet, where techniques to efficiently transfer quantum states between remote quantum processors distributed in the network play a central role. 
	
	In this review, we describe recent technological developments that rely on well-known laser physics and advanced modulation techniques to encode information for quantum cryptography with high fidelity and efficiency. This approach is compatible with a wide range of protocols and holds great potential for QKD implementations using photonic integrated chips.
	 
	\subsection{Quantum Bits and Information Encoding}
\begin{figure}[tb]
		\centering
		\includegraphics[width=0.85\columnwidth]{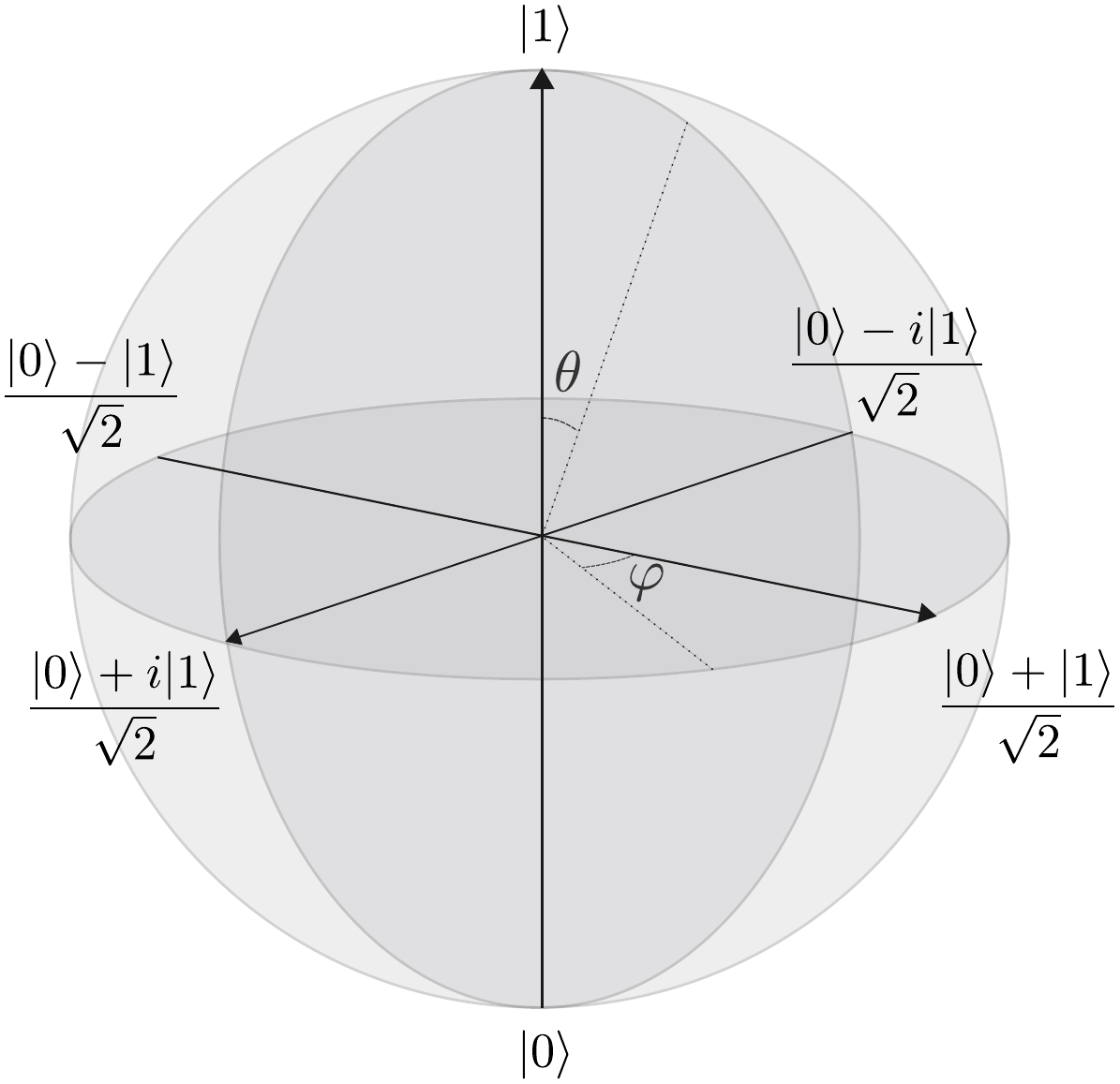}
		\caption{Bloch sphere representation of the pure states of a 2D Hilbert space.}
		\label{fig:intro:bloch:sphere}
	\end{figure}

	Encoding information in a quantum bit (qubit) is equivalent to engineering a superposition of two eigenstates, $|0\rangle$ and $|1\rangle$, of a two-level system. Qubits states are vectors of a 2-dimensional Hilbert space and are best represented in the Bloch sphere, with the eigenstates located at the poles of the sphere, as shown in Fig. \ref{fig:intro:bloch:sphere}. 
	Photons are the most versatile information carrier as they are easy to generate, manipulate, and transmit over long distance over free-space or fiber channels. Most importantly photons offer multiple degrees of freedom suitable to encode information as quantum bits. These include polarization, frequency, time and phase, orbital angular momentum, spatial mode etc. No matter which degree of freedom is selected, encoding the qubit state always comes down to encoding a state of the Bloch sphere, i.e. encoding 2 angles: a polar angle, $\theta$ that determines the coefficients of the superposition of the two eigenstates, and an azimuthal angle $\varphi$, that determines the phase of the superposition. The qubit state can be written as
	\begin{equation}
	    |\Psi\rangle = \cos{\frac{\theta}{2}} |0\rangle + e^{i\varphi}\sin{\frac{\theta}{2}} |1\rangle.
	\end{equation}
    
    In terms of Pauli matrices, the polar states $|0\rangle$ and $|1\rangle$ correspond to the eigenstates of $\sigma_Z$, while the eigenstates of $\sigma_X$ and $\sigma_Y$ are located on two perpendicular diameters of the equator. They correspond to an equal superposition of $|0\rangle$ and $|1\rangle$, with superposition phase $\varphi = 0, \pi$ and $\varphi = \pi/2, 3\pi/2$ for the $\sigma_X$ and $\sigma_Y$ eigenstates, respectively. Of course, there is an infinity of equatorial states. As detailed later, the ones we list are particularly interesting for quantum information and quantum cryptography because the Pauli matrices form a set of conjugate bases for the Hilbert space. Hence, the eigenstates of one basis are an equal superposition of the eigenstates of the other.
    
    In practice, if the two polar states can be accessed and prepared independently, then any qubit state can be engineered using intensity modulation to encode $\theta$ and phase modulation to encode $\varphi$.

	\subsection{Quantum Cryptography}
	The central problem of cryptography is the secure exchange of encryption keys between distant parties, Alice and Bob. While conventional cryptography protocols use computationally secure techniques to establish a key, quantum cryptography provides information-theoretic secure methods based on the exchange of quantum states, whereby a quantum statistical analysis provides a measure of the information leakage to a potential eavesdropper, Eve. 
	
\begin{figure}[tb]
		\centering
		\includegraphics[width=\columnwidth]{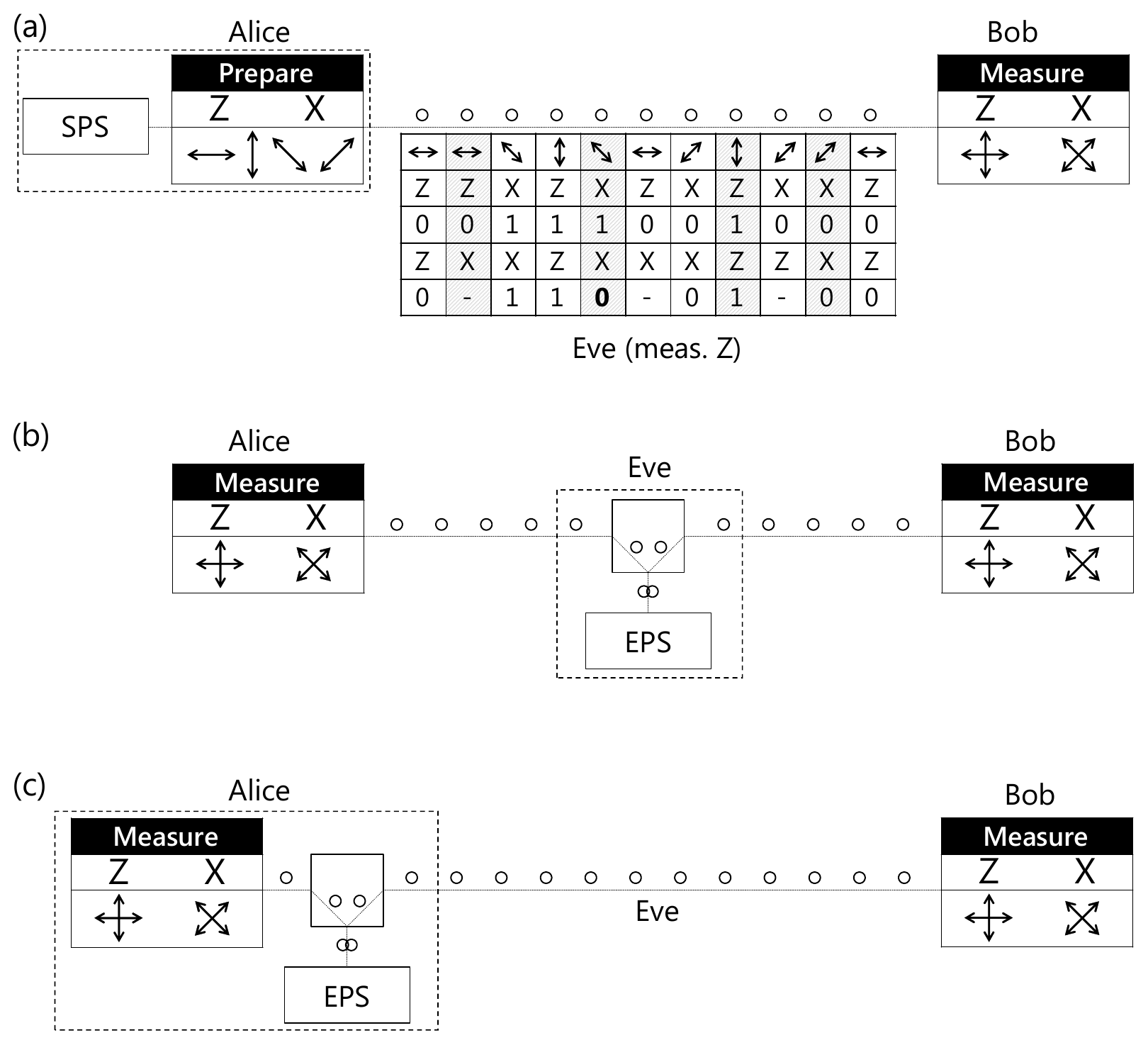}
		\caption{(a) Original BB84 QKD protocol and sifting. Alice prepares the polarization states of photons from a single photon source (SPS) by selecting randomly one of 2 conjugate polarization bases, $Z$ and $X$. Bob measures the polarization states of incoming photons by randomly selecting the measurement basis. In the sifting procedure, Alice and Bob compare their bases and discard the events where the bases do not match. Eve performs an intercept-and-resend attack on some of the transmitted photons (greyed columns), by measuring the $Z$ basis. Eve's attack generates a bit flip error in the $X$ basis with 50~$\%$ probability. By measuring the error rate in the sifted keys, Alice and Bob can quantify the amount of information gained by Eve. (b) E91 protocol. Both Alice and Bob measure one half of an entangled photon pair emitted from a source located in the middle. After sifting, Alice and Bob confirm the violation of Bell's inequalities on a subset of the sifted keys. The protocol is device independent. Even if Eve controls the source she cannot tamper with it without affecting the violation of Bell's inequalities and thus be detected. (c) Equivalence between the BB84 protocol and the E91 protocol with the entangled photon pair source (EPS) is located at Alice.}
		\label{fig:intro:bb84}
	\end{figure}

	\paragraph{The BB84 protocol}
	For the sake of didactic, we illustrate our discussion based on the first quantum key distribution (QKD) protocol, devised by Bennet and Brassard in 1984 \cite{Bennett1984}. The BB84 protocol originally proposed to encode information using two orthogonal polarization states of single photons, and along two conjugate bases, such that an eigenstate in one basis corresponds to an equal superposition of both eigenstates of the other. The protocol, illustrated in Fig. \ref{fig:intro:bb84}, operates as follows: 
\begin{enumerate}
    \item{\emph{Prepare.}} Alice prepares a stream of photons in different polarization eigenstates of one or the other of the two bases, all selected at random for each photon, and sends it to Bob. 
    \item{\emph{ Measure.}} Bob measures the incoming photons in one of the two bases, again selected randomly for each photon.
    \item{\emph{Sifting.}} After the photon transmission is complete, Alice and Bob proceed to the sifting procedure, during which they compare their bases choices and only keep the events where the `prepare' and `measure' bases match. In each basis, one of the eigenstates is attributed for the logical bit $0$ and the other to the logical bit $1$.
    \item{\emph{Error Correction.}} In the ideal case, Alice and Bob are then left with the same sequence of bits. In practice, Alice and Bob's sifted keys would slightly differ and need to be error-corrected before being used as cryptographic keys.
    \item{\emph{Privacy Amplification.}} The key point that ensures the security of the protocol is that an eavesdropper attempting to measure the transmitted photons between Alice and Bob would necessarily generate errors in the keys. Because of the no-cloning theorem, Eve cannot duplicate an unknown incoming quantum state. Since Alice and Bob select their bases randomly, Eve can only `guess' which basis to measure and resend a photon in the state she measured. If Eve's basis differs from Bob's, then an error is introduced with 50~$\%$ probability. By monitoring the error rate, Alice and Bob can therefore infer how much knowledge Eve gained about the key and use a privacy amplification algorithm to reduce this knowledge to a negligible amount.
\end{enumerate}

    \paragraph{The E91 protocol}
    An alternative protocol, based on the measurement of non-local correlation inherent to entangled photon pairs was devised in 1991 by Ekert \cite{Ekert.1991}. In this protocol, termed E91 or entanglement distillation protocol, Alice and Bob each receive one photon from a polarization entangled photon pair and measure them in randomly selected bases. In the sifting procedure Alice and Bob retain the detection events where their basis choices match and attribute the logical bits accordingly. In order to certify the security of the channel, Alice and Bob confirm the violation of Bells inequalities on a subset of randomly selected and uniformly distributed detection events. Non-locality ensures that the protocol is inherently device independent and information-theoretically secure. Even if the source of entangled photon were controlled by Eve, it would be impossible to predict which bases will be used for measurement or to measure the photon without destroying the correlations. The information-theoretic security of the BB84 protocol was later established by proving the equivalence of the BB84 protocol with the E91 protocol, where the entangled photon pair source is located at Alice (see Fig. \ref{fig:intro:bb84}~b and c) \cite{Gottesman.2004}. 
	
	\subsection{Sub-Poissonian Light Sources}
	
	Implementing the BB84 or the E91 protocols as they were proposed originally requires efficient sources of single photons or entangled photon pairs. 
    In practice, such sources always present non-idealities so it is challenging to guarantee that they emit exactly one photon (or one photon pair) at a time. The quality of these sources therefore is evaluated in terms of their single photon purity, their emission rate (or brightness) and the indistinguishability of the emitted photons \cite{Senellart.2017}. 
    Single photon sources (SPS) are characterized by their sub-Poissonian statistics typically evidenced as an anti-bunching in the second-order intensity correlation function $g^{(2)}(0)<1$, as measured in a typical Hanbury Brown and Twiss experiment.  The value of $g^{(2)}(0)$ is used to quantify the purity of the source.
    
    In the last 2 decades significant progress has been made to develop bright, on-demand single photon sources. 
    Spontaneous parametric down conversion (SPDC) sources naturally produce entangled photon pairs as required for the E91 protocol \cite{Kwiat.1995}, and can be used for heralded generation of single photons where the detection of one photon of the pair heralds the presence of the counterpart single photon. There is however a trade-off between purity, indistinguishability and brightness for these sources, as the former two decrease as the latter increases \cite{Senellart.2017}. SPDC was used to demonstrate entanglement-based QKD over a 144 km free-space link in 2007 \cite{Ursin.2007}, and in 2020 over 1,120 km in a satellite-to-ground experiment \cite{Yin.2020}.
    
    The best performing SPS to date are semiconductor quantum dots (QD), with which purity, brightness and indistinguishability can all be maximized at the same time \cite{Wang.2019,Hanschke.2018,Kupko.2020}. 
     QD sources were used to demonstrate the BB84 protocol in an early 2000 experiment with a device with $g^{(2)}(0)=0.14$\cite{Waks.2002}. In 2015, the performance was improved in a new demonstration with source purity $g^{(2)}(0)=0.002$, achieving a QKD range of 120~km~\cite{Takemoto.2015}.
    QD sources can also be used to generate single photon pairs via the biexcitonic decay, and were used more recently to demonstrate entanglement-based QKD \cite{BassoBasset.2021,Schimpf.2021} at higher rates than with SPDC sources.

	\subsection{Weak Coherent Pulses}
	Because of the challenges in realizing efficient on-demand  single photon sources, a more practical approach based on weak coherent laser pulses was introduced in the early 90’s \cite{CharlesH.Bennett.1992,Huttner.1995}. 
	 Weak coherent pulses (WCPs) present the advantage of being simple to generate, in particular at much higher repetition rates. 
	 While QKD with deterministic SPS was demonstrated with clock rates up to the 100~MHz range, WCPs-based QKD has been demonstrated at clock rates of 5~GHz for the BB84 protocol \cite{Grunenfelder2020} and up to 10~GHz for the differential phase shift protocol \cite{Choi.2010,Takesue.2007}. This allows for secure key rates several orders of magnitude higher than currently achievable with SPS, with QKD up to 13~Mb/s demonstrated in a WCP QKD system~\cite{Yuan.2018}.
	
	In addition, the lasers operate at room temperature, which avoids the need of cryogenic equipment. Finally, WCPs also provide a convenient way to implement protocols using other degrees of freedom than polarization, as for example frequency, time-bin, or orbital angular momentum. 
	In order to be suitable for QKD, the pulses are required to be identical for any observable not used for information encoding. In other words, the observable used to encode the quantum state should be completely decoupled from all the other observables. Any information correlation between encoding and non-encoding observables would introduce a side-channel and allow an eavesdropper to gain information on the encoded state without being detected.
	
	The main inconvenience of WCPs comes from the multi-photon contributions, that open a vulnerability to information leakage \cite{Huttner.1995,Brassard.2000,NorbertLutkenhaus.2000}. 
	
	A coherent state $|\alpha\rangle$ of mean photon number $\mu$ per unit time and phase $\theta$ formally reads
	\begin{equation}\label{eq:coherent:state} 
	   |\alpha\rangle = e^{-\mu/2} \sum_{n=0}^{\infty} \frac{(\sqrt{\mu} e^{i\theta})^n}{\sqrt{n!}} |n\rangle .
	\end{equation}
	
	At each unit time, the outcome of a photon number measurement from a source of coherent states of mean photon number $\mu$ per unit time is $n$ photons with a probability given by the Poisson distribution
	\begin{equation}\label{eq:poissonian}
	    P(n;\mu) = \frac{e^{-\mu} \mu^n}{n!}.
	\end{equation}
	Beside the fact that excess photons could leak from a pulse during its propagation, a well-engineered photon number splitting (PNS) attack could be designed to extract and store one excess photon, thus allowing the eavesdropper to access information in the key without being detected \cite{NorbertLutkenhaus.2000}.
	
	\begin{figure}[tb]
		\centering
		\includegraphics[width=\columnwidth]{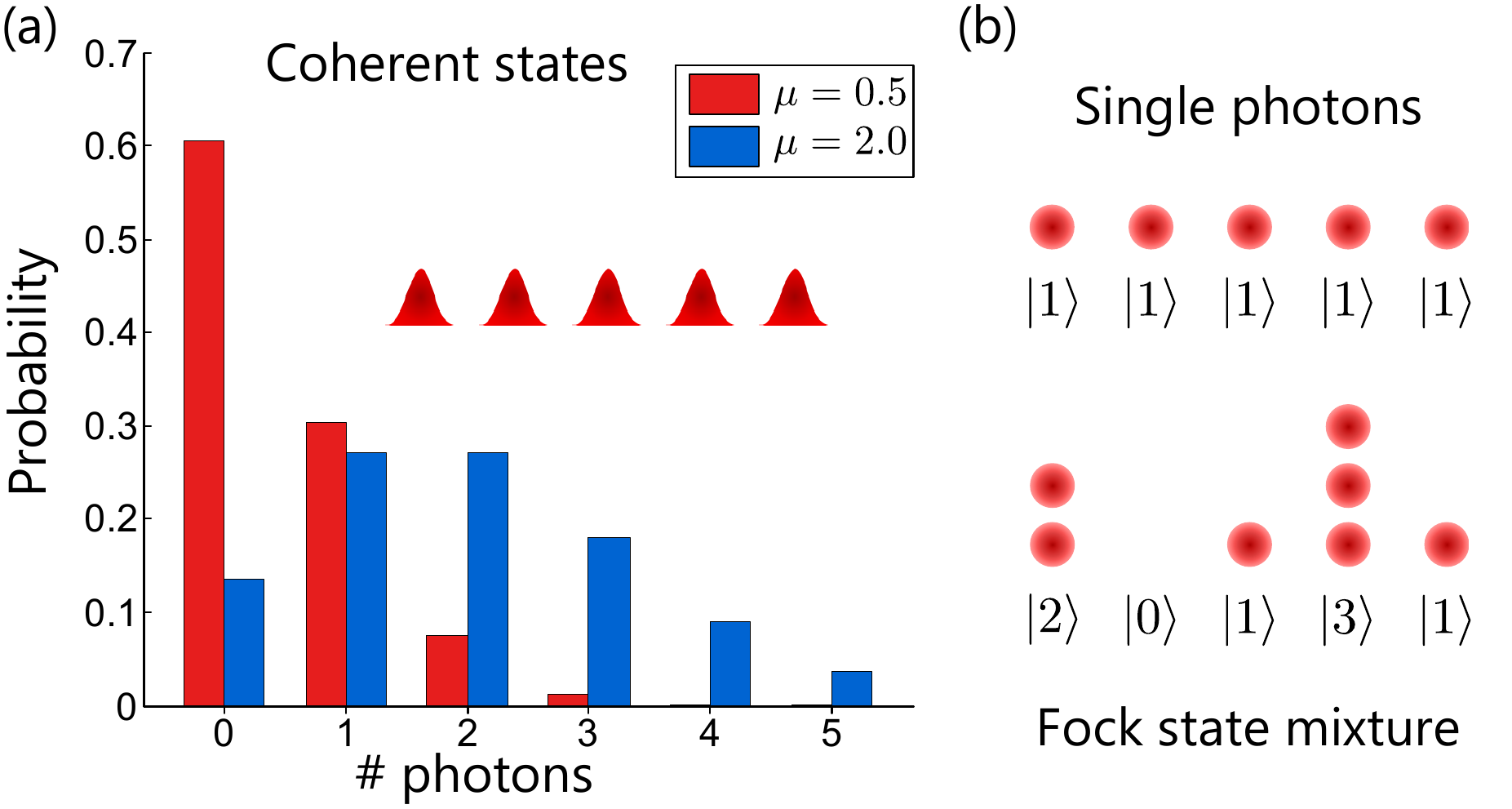}
		\caption{a. Photon number distribution in weak coherent pulses. The multi-photon contribution is suppressed by setting the mean photon number to lower than 1 b. In the presence of phase-randomizing the description of weak coherent signal pulses is equivalent to that of a mixture of photon number eigenstates (Fock states). Compared to an ideal source of single photon pulses there is a non-zero probability to emit multi-photon pulses.}
		\label{fig:intro:wcp}
	\end{figure}

    To ensure that the contribution of the multi-photon pulses is suppressed compared that of the single photon pulses, $\mu$ should be set to much lower than 1. This is illustrated in the example shown in Fig. \ref{fig:intro:wcp}~a: for the $\mu=2$ Poisson distribution, the 1- and 2-photon terms have equal contributions, while for $\mu=0.5$, the 1-photon term dominates the non-empty terms, with $P(1;0.5) = 4 \times P(2;0.5) = 24 \times P(3;0.5)$. 

\begin{figure}[tb]
		\centering
		\includegraphics[width=0.9\columnwidth]{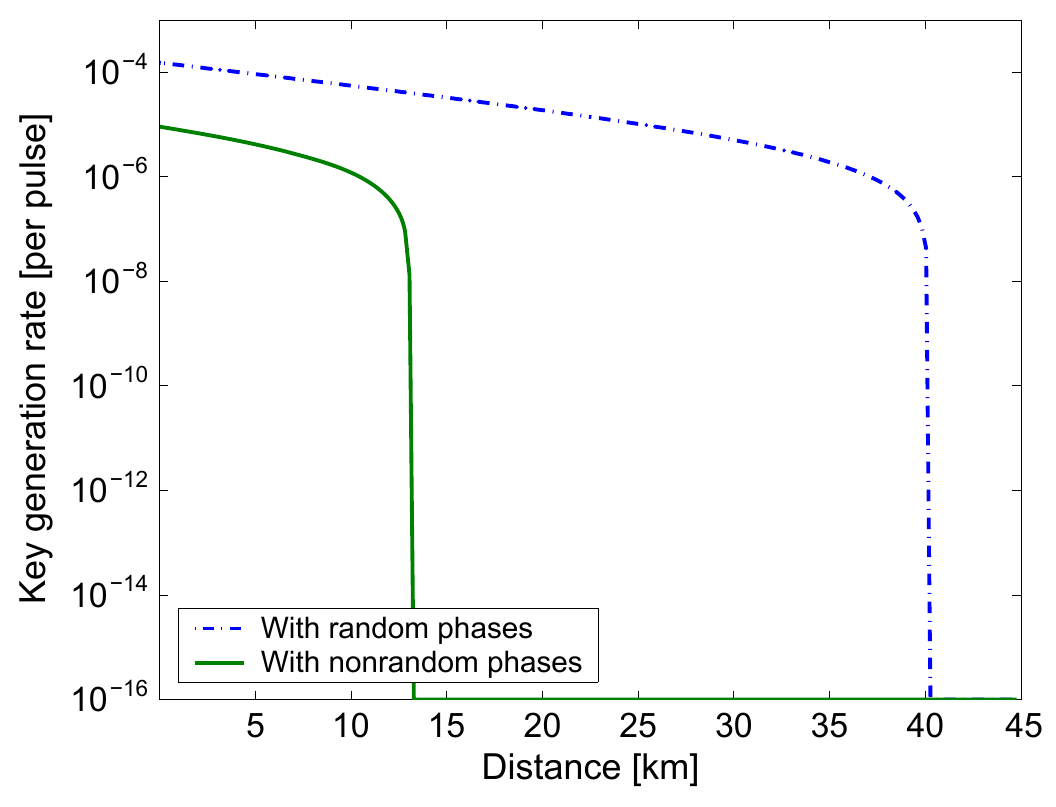}
		\caption{Security of the BB84 protocol with weak coherent pulses in the presence/absence of phase randomization. (\emph{Reproduced with permission from Ref.~\cite{Lo.2007}  \copyright{} Rinton Press.})}
		\label{fig:intro:phase:rand}
	\end{figure}

    A powerful practice to enhance the security is to randomize the phase of each signal emitted by the coherent source \cite{Brassard.2000, NorbertLutkenhaus.2000,Lo.2007}. In that case, the density matrix of the emitted state becomes
	\begin{equation}\label{eq:density:matrix}
	    \rho_\mu = \int_0^{2\pi} \frac{d\theta}{2\pi}|\alpha\rangle \langle \alpha| = e^{-\mu} \sum_{n=0}^{\infty} \frac{\mu^n}{n!} |n\rangle \langle n|
    \end{equation}
	i.e. indistinguishable from the density matrix of a mixture of Fock states. As illustrated in Fig. \ref{fig:intro:wcp}~b, for an eavesdropper having no knowledge of the phase, it would be as if Alice actively prepared and emitted Fock states with probabilities taken from the Poisson distribution $P(n;\mu)$. By treating each signal pulse as a photon number eigenstate, the security analysis can be drafted for the most general attacks and the protocol performance, in particular the distance at which a secret key can be generated, is thereby greatly enhanced compared to the case where all signals are weak coherent pulses sharing the same phase reference (see Fig. \ref{fig:intro:phase:rand}) \cite{Lo.2007}.

	\subsection{Outline}
	Deploying QKD, and more generally quantum communication technologies, requires  solutions to encode quantum states in light pulses in an efficient and scalable way. While weak coherent pulses can readily be produced at very high rate from an attenuated train of bright laser pulses, there are essential requirements to be met by such sources to be practical for quantum applications. For quantum statistics to play a significant role, the pulses must be highly phase-coherent, with low intensity, phase, and frequency noise. In particular, the pulses must be highly indistinguishable - except in the degree of freedom used to encode quantum information.

	This review article presents the theory and applications of a recent framework based on advanced laser technology developed precisely for the purpose of generating such high quality pulses. This framework, termed phase-seeding, exploits gain-switching, direct phase modulation and optical injection locking and was successfully applied to various QKD protocols, although its scope is much broader.
	
\begin{figure}[tb]
		\centering
		\includegraphics[width=1\columnwidth]{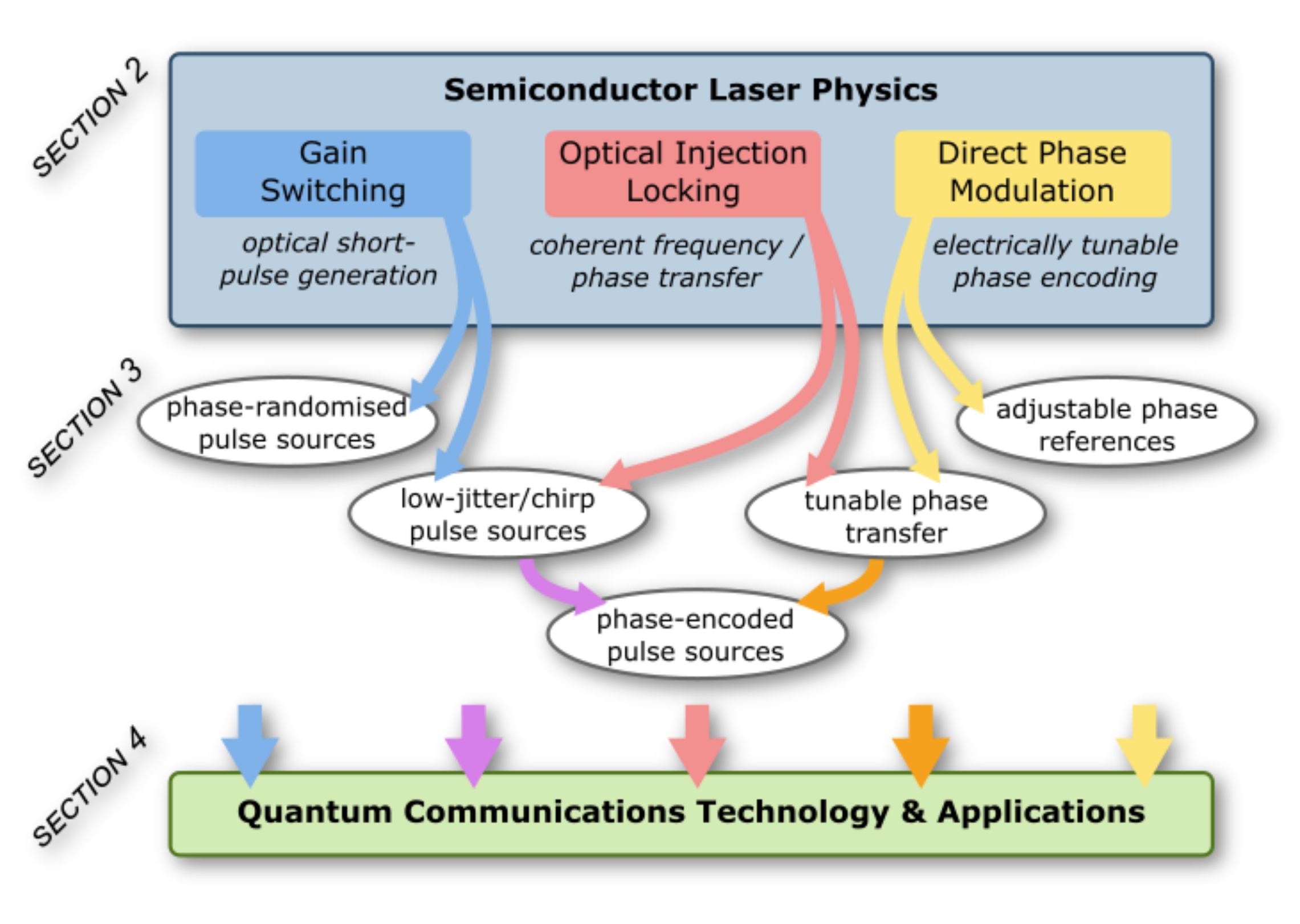}
		\caption{Visual outline of review article, highlighting the exploitation of concepts in semiconductor laser physics to develop new quantum communications technologies and applications.}
		\label{fig:intro:summary}
	\end{figure}

	Figure \ref{fig:intro:summary} is an outline of the topics discussed in this review and a summary of the main acronyms used is presented in Table I.
	Section \ref{sec:theory} introduces the theory of laser physics and optical injection locking (OIL) in the presence of noise through a rate equation model. It is shown how gain switching provides a source of phase randomized pulses, how stimulated emission from an injected seed can be used to suppress noise in the pulses, and how direct modulation can be used to induce deterministic phase shifts.
	
	Section \ref{sec:direct:modulated:sources} dives more into the phenomenology accessible with direct modulated light sources to show how judiciously combining the different laser properties can make these sources suitable for quantum communications and beyond. In particular it is shown how a phase randomized pulse source can be locked to a tunable-phase optical seed to yield an efficient, versatile and practical source of phase-encoded pulses.
	
	Experimental demonstrations and applications to quantum random number generation, coherent optical communications and to the most common quantum key distribution protocols are reviewed in Section \ref{sec:applications}. 
	
	To conclude, Section \ref{sec:discussion} provides a discussion and general perspectives on the future potential of phase-seeding.

		\begin{table}
		\caption{\label{tab:acronyms} Main acronyms used in this review.}
	\begin{tabular}[htbp]{@{}ll@{}}
		\hline
		Acronym & Meaning\\ 
		\hline
ADC	&	Analogue-to-digital converter	\\
BB84 & Bennet and Brassard protocol (1984) \\
COW	&	Coherent-one-way	\\
CV	&	Continuous-variable 	\\
CW	&	Continuous-wave	\\
DFB	&	Distributed feedback	\\
DPR	&	Distributed phase reference	\\
DPS	&	Differential phase shift	\\
DPSK	&	Differential phase shift keying	\\
DQPS	&	Differential quadrature phase-shift \\
DQPSK	&	Differential quadrature phase-shift keying	\\
$\ \ $  RZ-DPSK	&	Return-to-zero DPSK	\\
DV	&	Discrete-variable	\\
E91 & Ekert protocol (1991) \\
EOPM	&	Electro-optic phase modulator	\\
EPS	&	Entangled photon pair source	\\
GS	&	Gain switching	\\
HOM	&	Hong-Ou-Mandel	\\
IQ plane	&	In-phase-Quadrature plane	\\
MDI	&	Measurement device independent	\\
MZI	&	Mach-Zehnder interferometer	\\
$\ \ $  AMZI	&	Asymmetric Mach-Zehnder interferometer	\\
MZM 	&	Mach-Zehnder modulator	\\
OIL	&	Optical injection locking	\\
PD	&	Photodiode	\\
PNS 	&	Photon number splitting attack	\\
QAM	&	Quadrature amplitude modulation	\\
QBER	&	Quantum bit error rate	\\
QD 	&	Quantum dot	\\
QKD	&	Quantum key distribution	\\
QRNG	&	Quantum random number generator	\\
RIN	&	Relative intensity noise	\\
SKR	&	Secure key rate	\\
SPDC	&	Spontaneous
parametric down conversion	\\
SPS	&	Single photon source	\\
TF-QKD	&	Twin-field QKD	\\
VCSEL	&	Vertical-cavity surface-emitting laser	\\
WCP	&	Weak coherent pulse	\\
		\hline
	\end{tabular}
\end{table}

	\section{Theory}
	\label{sec:theory}
	
	Lasers are complex dynamical systems, which can exhibit a broad range of operating states with diverse temporal and spectral properties.
	An even greater parameter space of optical waveforms can be generated through the interaction of multiple such devices.
	To harness this flexibility, however, precise control of the driving conditions and the interaction between coupled devices is required, necessitating a detailed understanding of the underlying laser dynamics.
	Therefore, we begin by introducing the fundamental laser physics which underpin directly amplitude and phase modulated light source technologies, in addition to presenting a rate equation model which can be used to accurately simulate the relevant laser dynamics.

	\subsection{Rate Equation Model}
	Laser diodes are widely used for fiber-optic communications and have benefited from decades of research into their characteristics. This has resulted not only in low-cost, high-performance devices, but also in a strong theoretical understanding of their behavior. In particular, the widely accepted method for modeling semiconductor lasers is using rate equations~\cite{Agrawal1993,Srinivasan1997a,Troger1999,Ahmed2001,Fatadin2006}. 
	These describe the interactions between three relevant quantities in laser dynamics: the carrier density $N$, the photon density $S$ and optical phase $\phi$.
	In the following, we introduce the rate equations and apply this well-established technique to describe and simulate laser effects which can be exploited for encoding quantum information.
	
	The rate equations for a single-mode laser cavity are~\cite{Srinivasan1997a,Troger1999}: 
	\begin{align}
		\frac{d N(t)}{d t} &=\frac{I(t)}{qV}-\frac{N(t)}{\tau_{n}}-g \frac{N(t)-N_{0}}{1+\epsilon S(t)} S(t) + F_N(t) \label{eq:N} \\
		\frac{d S(t)}{d t} &=\Gamma g \frac{N(t)-N_{0}}{1+\epsilon S(t)} S(t)-\frac{S(t)}{\tau_{p}}+\frac{\Gamma\beta N(t)}{\tau_{n}} + F_S(t) \label{eq:S} \\
		\frac{d \phi(t)}{d t} &=\frac{\alpha}{2} \left[ \Gamma g(N(t)-N_0) -\frac{1}{\tau_p}\right] + F_\phi(t) \label{eq:phi}
	\end{align}
	where $I(t)$ is the applied current, $q$ is the electron charge and $V$ is the active layer volume. $\tau_n$ and $\tau_p$ are the carrier and photon lifetimes respectively, which quantify the average time a carrier or photon survives in the laser cavity. $\Gamma$ is the mode confinement factor which accounts for the fact that only a fraction $\Gamma$ of the photons are confined to the active layer. $g$ is the differential gain coefficient which arises from making the approximation that the gain is linear as a function of carrier density. $\epsilon$ is the gain compression factor that accounts for the non-linear reduction in gain at high power outputs \cite{koch_effect_1986}. $N_0$ is the carrier density at transparency and $\beta$ is the fraction of spontaneous emission coupled into the lasing mode. Finally, $\alpha$ is the linewidth enhancement factor which quantifies the increase in linewidth due to the coupling between refractive index and carrier density in semiconductor lasers \cite{henry_theory_1982}. The power output of the laser is related to the photon density by:
	\begin{equation} \label{eq:power}
		P(t)=\frac{V \eta h \nu}{2\Gamma \tau_{p}} S(t)
	\end{equation}
	where $\eta$ is the differential quantum efficiency, $h$ is Planck's constant and $\nu$ is the laser frequency. The rate equation parameters are intrinsic properties of each laser and will vary from laser to laser. There are various experimental methods that can be used to obtain estimates for them \cite{bjerkan_measurement_1996, cartledge_extraction_1997}.
 
	The terms $F_N$, $F_S$, and $F_{\phi}$ are so-called Langevin noise terms. 
These take on different forms depending on the sources of noise that are being considered.
	Accounting for the effects of spontaneous emission, they are given by:
	\begin{align}
		F_S(t) &= \sqrt{\frac{2\Gamma \beta N(t) S(t) }{\tau_n \Delta t}} \cdot x_S \\
		F_\phi (t) &= \sqrt{\frac{\Gamma \beta N(t)}{2 \tau_n S(t) \Delta t}} \cdot x_\phi \\
		F_Z (t) &= \sqrt{\frac{2N(t)}{V\tau_n \Delta t }} \cdot x_Z \\
		F_N (t) &= F_Z(t) - \frac{F_S(t)}{\Gamma}
	\end{align} where $F_Z(t)$ is a noise term, uncorrelated to $F_S(t)$ and $F_\phi (t)$, used to define the carrier density noise term $F_N(t)$. $\Delta t$ is the integration time step and $x_S$, $x_\phi$ and $x_Z$ are three independent standard normal random variables. Often the rate equations are used without noise terms, when the effects of noise are not of interest. In this case the rate equations can be solved using standard numerical integration tools. However, when the noise terms are included, the rate equations become stochastic differential equations and must be solved using stochastic numerical integration methods, the simplest of which is the Euler-Maruyama method \cite{higham_algorithmic_2001}.

It should be noted that various forms of semiconductor laser rate equations can be found in literature, using different complexity of models to simulate physical phenomena and occasionally using units of photon/carrier number rather than density.
Therefore, care must be taken when selecting appropriate equations and parameters from the literature.
Here, we base our simulations on parameters from Ref.~\cite{Srinivasan1997a}, obtained by fitting parameters to DFB laser experiments---as summarized in Table \ref{tab:rate_equation_parameters}. 

	\begin{table}
		\caption{\label{tab:rate_equation_parameters} Typical rate equation parameters (after Refs. \cite{cartledge_extraction_1997, troger_novel_1999}).}
	\begin{tabular}[htbp]{@{}lll@{}}
		\hline
		Parameters & Values & Description\\ 
		\hline
		$\tau_n$ (ns) & 0.74 & Carrier lifetime\\
		$\tau_p$ (ps) & 0.74 & Photon lifetime \\
		$g$ ($\times 10^{-6}~\textrm{cm}^3\textrm{s}^{-1}$) & 1.27 & Differential gain coefficient \\
		$\varepsilon$ ($\times 10^{-17}~\textrm{cm}^3$) & 1.18 & Gain compression factor \\
		$N_0$ ($\times 10^{18}~\textrm{cm}^{-3}$) & 0.85 & Carrier density at transparency \\
		$\beta$ ($\times 10^{-5}$)& 0.50 & Spontaneous emission factor\\
		$\alpha$ & 2.7 & Linewidth enhancement factor\\
		$\eta$ & 0.20 & Differential quantum efficiency\\
		$V$ ($\times10^{-11}~\textrm{cm}^3$) & 1.72 & Active layer volume\\
		$\Gamma$ & 0.27 & Mode confinement factor \\
		$\kappa$ ($\times 10^{11}~\textrm{Hz}$) & 1.13 & OIL coupling term\\
		\hline
	\end{tabular}
\end{table}

By numerically solving the rate equations, the time-dependent laser output power $P(t)$ and phase $\phi(t)$ can be obtained for any given current input function $I(t)$. 
Using this model we will next elucidate the effects of gain switching, direct phase modulation and optical injection locking.

\subsection{Gain Switching (GS)}
\begin{figure}[tbph]
	\centering
	\includegraphics[width=\columnwidth]{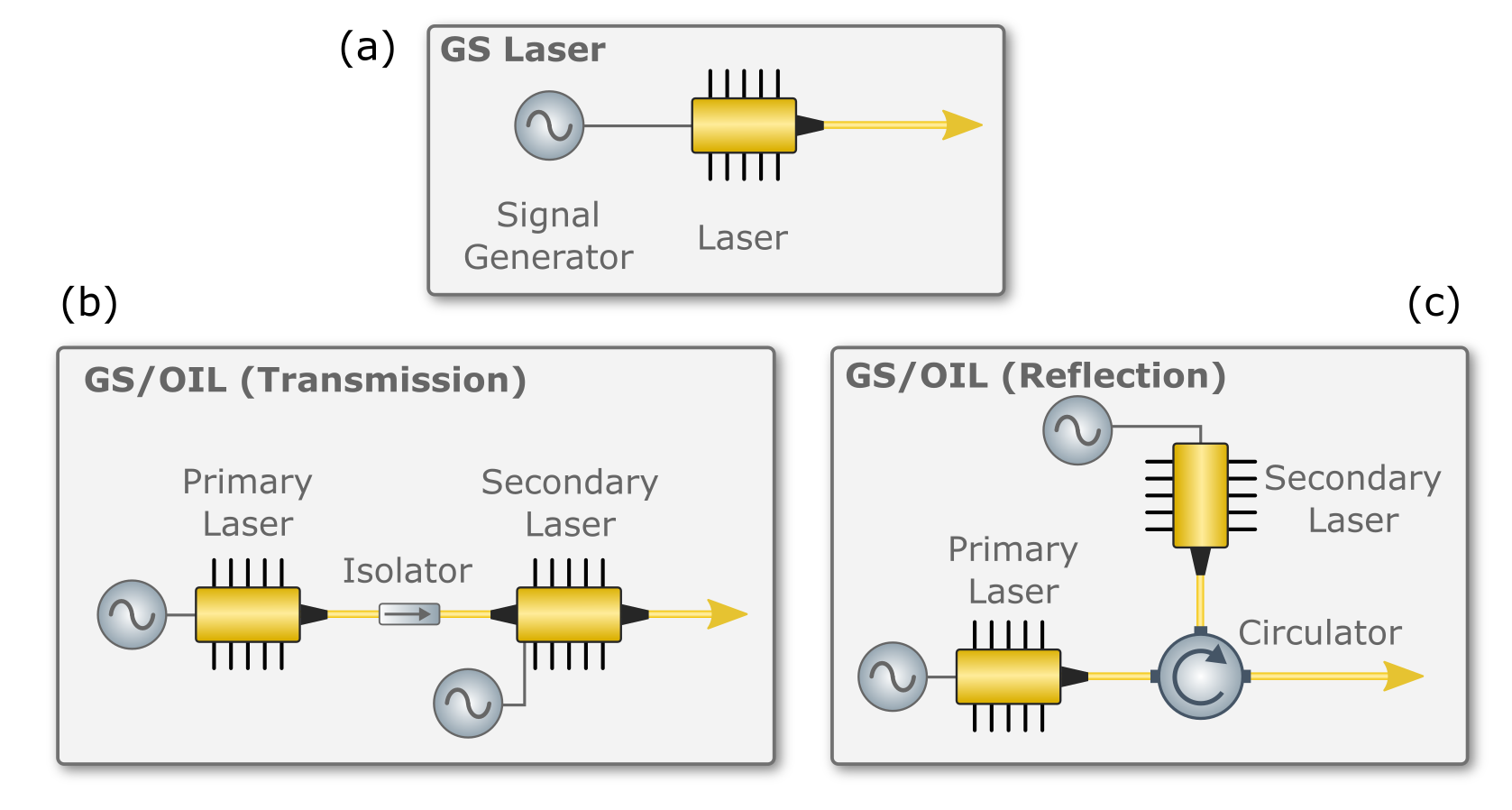}
	\caption{Directly modulated laser setups: (a) gain-switched laser; gain-switched optical injection locked systems in (b) transmissive and (c) reflective configuration.}
	\label{fig:setup}
\end{figure}

Gain switching is a widely used approach for optical pulse generation via large-signal modulation of the electrical pump power, periodically driving the laser above and below the lasing threshold to cause periodic emission of light~\cite{Siegman1986}.
This benefits from a simple, compact experimental setup, comprising only an electrical signal generator connected to a laser source, which is herein assumed to be a semiconductor laser diode (Fig~\ref{fig:setup}(a)).
The optical output does not simply follow the shape of the electrical signal, however.
Instead, the temporal properties of optical emission depend on interplay between photons and electrically-injected carriers in the laser cavity, where carriers in semiconductor lasers typically have nanosecond-duration lifetimes.

In the steady-state, cavity gain balances loss and the carrier density is clamped at the carrier density threshold, but when a large current is first applied to the laser, carriers build up quickly and can temporarily overshoot the carrier density threshold.
This results in a large emission of photons which subsequently deplete the carriers.
The resulting interplay between photons and carriers causes damped oscillations (and corresponding phase fluctuations), known as relaxation oscillations.

\begin{figure}[tb]
	\centering
	\includegraphics[width=0.95\columnwidth]{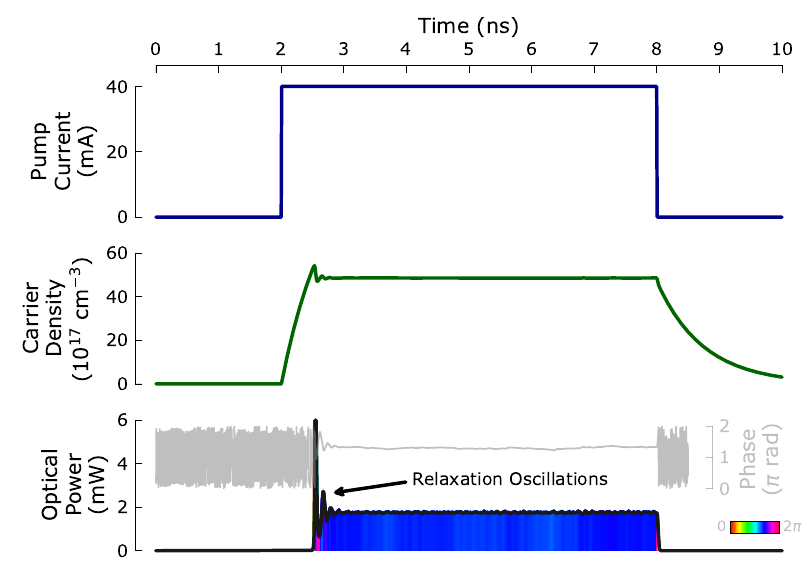}
	\caption{Simulated laser pulse generation through gain-switching, highlighting relaxation oscillation and slow carrier decay phenomena.}
\label{fig:gs}
\end{figure}

If the electrical drive pulse is short, photon emission can be extinguished after the first oscillation, forming a short Gaussian-shaped optical pulse with duration on the order of 10s~ps.
Such gain-switched pulses often also have a frequency chirp across the pulse.
By sustaining the electrical current for longer, however, relaxation oscillations continue, gradually being damped to reach the steady-state and forming a characteristic rectangular-shaped pulse with an initial overshoot (as shown in Fig.~\ref{fig:gs}).
When the pump current is removed, carrier density decreases---there is initially a sharp fall as stimulated emission rapidly depletes carriers that are no longer being replenished, which causes the gain to fall below threshold and inhibits lasing.
Even though the optical power falls sharply at this point, a gradual decay of the remaining cavity carrier density continues through the slower spontaneous emission process (note the tail-off in carrier density in Fig.~\ref{fig:gs}).
Both the short Gaussian pulse and rectangular-shaped pulse emission formats can be usefully exploited for communication applications.

The relaxation oscillation frequency defines the maximum pulse generation rate and the shortest possible pulse duration, which are very important factors for applications.
This frequency is affected by various factors including the gain medium upper-state lifetime, laser geometry and driving conditions.
As semiconductor lasers typically have short upper-state lifetimes ($\sim$nanoseconds), high-performance GHz repetition rate gain-switched pulse generation is possible, which makes them very attractive for optical communications.

\begin{figure}[tb]
\centering
\includegraphics[width=0.95\columnwidth]{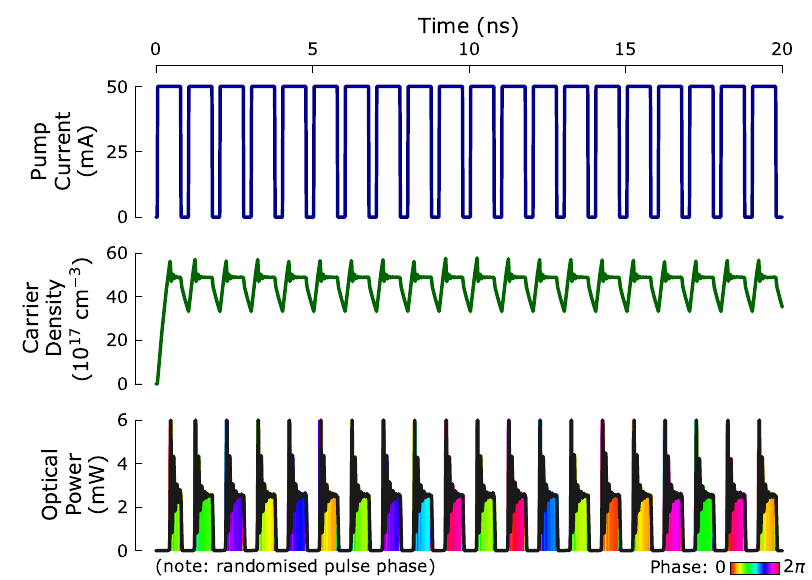}
\caption{Simulated gain-switching dynamics showing the generating of a phase-randomized pulse train.}
\label{fig:gs_rand}
\end{figure}

Another important aspect of gain-switching is that the stimulated emission process amplifies photons in the laser cavity, where this `seed light' is provided by spontaneous emission if the laser is initially off.
As the phase of spontaneous emission is intrinsically random (seeded by vacuum fluctuations), the steady-state phase for gain-switched pulses is effectively random, providing the light build-up starts from a near-empty cavity. 
Therefore, with periodic application of the pump current, a repetitive pulse train can be produced, where each pulse has a random steady-state phase, as shown in Fig.~\ref{fig:gs_rand}.

\subsection{Optical Injection Locking (OIL)} \label{sec:OIL}

We now consider an arrangement comprising two lasers in a primary / secondary (i.e. master / slave) configuration, where light from the primary laser is injected into the cavity of the secondary laser.
Injected light can coherently seed the stimulated emission process and under appropriate conditions, can thus fully determine the wavelength and phase of the generated secondary laser emission.
This can be implemented in a linear arrangement, if the secondary laser has partially reflective facets on both sides of the cavity (Fig.~\ref{fig:setup}b), or alternatively using a circulator (Fig.~\ref{fig:setup}c).
In both cases, the primary laser is isolated from reflections coming back from the secondary laser.

There are also benefits to injection locking on the pulse performance of the secondary laser.
As pulses are deterministically seeded by injection, rather than by random vacuum fluctuations, the temporal jitter, relative intensity noise (RIN) and pulse chirp can all be significantly reduced~\cite{Lau2009,Liu2020b}.
Injection also increases the relaxation oscillation frequency, causing faster damping of transients and smaller overshooting from the steady state, as well as enhancing the modulation bandwidth compared to the same laser in free-running operation.

To model the effects of OIL on the secondary laser, the rate equations can be extended by adding terms to the rate equations for $S$ and $\phi$\cite{lau_enhanced_2009}:

\begin{align}
\frac{dN(t)}{dt} &= \frac{dN_\textrm{fr}(t)}{dt} \label{eq:N_OIL} \\
\frac{dS(t)}{dt} &= \frac{dS_\textrm{fr}(t)}{dt} +2 \kappa \sqrt{S_{\text {inj }}(t) S(t)} \cos(\Delta\phi(t) - \Delta\omega\inj t) \label{eq:S_OIL} \\
\frac{d\phi(t)}{dt} &= \frac{d\phi_\textrm{fr}(t)}{dt} -\kappa \sqrt{\frac{S_{\mathrm{inj}}(t)}{S(t)}} \sin (\Delta\phi(t) -\Delta \omega\inj t)  \label{eq:phi_OIL}
\end{align}
where the subscript $_\textrm{fr}$ denotes the standard rate equations for a free running laser given by Eqns. \ref{eq:N}-\ref{eq:phi}. $\Delta\phi = \phi(t) - \phi\inj(t)$ is the difference between the secondary laser phase and the phase of the injected light,
$\kappa$ is a coupling coefficient which quantifies the rate at which injected photons enter the secondary laser cavity, $S\inj$ is the injected photon density and $\Delta \omega\inj$ is the difference in free-running optical angular frequency between primary and secondary lasers.
We note that Eqn.~\ref{eq:phi_OIL} describes the phase of the laser into which light is injected, although OIL rate equations are also occasionally expressed in the literature in terms of the phase difference between that laser's emitted light and the injected light---in this case, the frequency detuning $\Delta\omega_\mathrm{inj}$ appears outside the sine term and without a dependency on $t$, which can simplify the numerical solving algorithm~\cite{Liu2020b}.

The locking effect of optical injection can be seen by considering the sine term in Eqn. \ref{eq:phi_OIL}. 
For zero detuning, when the phase difference $\phi - \phi\inj$ is positive (in the interval $[-\pi,\pi)$), the term will be negative and vice versa. 
This has the effect of locking the secondary laser phase to the primary laser (up to a fixed phase offset).
However, a non-zero detuning $\Delta \omega\inj$ acts against this locking effect.
Stable OIL can therefore only be obtained provided the detuning falls within the ``locking bandwidth'', which can be derived from the steady state solutions of the OIL rate equations \cite{Liu2020b}:
\begin{equation}
-\kappa \sqrt{1+\alpha^{2}} \sqrt{\frac{P\inj}{P_0}} <\Delta \omega\inj< \kappa \sqrt{\frac{P\inj}{P_0}}
\end{equation}
where $P\inj$ is the injected optical power and $P_0$ is the free-running secondary laser power. 
The injection ratio $P\inj / P_0$ and primary-secondary detuning $\Delta \omega\inj$ are crucial parameters in OIL dynamics. 
Even when an OIL system is within the stable locking range, the exact impact upon modulation bandwidth and pulse performance depends strongly on both detuning and injection ratio, thus requiring careful design of OIL systems for each target application. 

\begin{figure}[tbph]
\centering
\includegraphics[width=0.95\columnwidth]{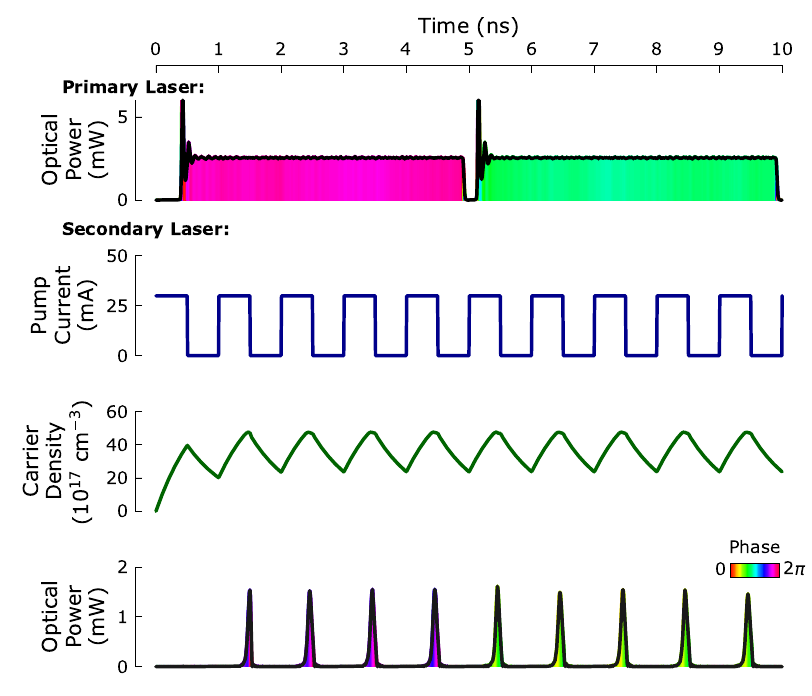}
\caption{Simulated generation of gain-switched pulse train, seeded by a long-pulsed primary laser. The secondary laser emits pulses that have a fixed phase relationship to the primary pulse phase (as illustrated using color).}
\label{fig:oil}
\end{figure}

To model the effects of OIL in a GS primary-secondary laser setup, two sets of rate equations must be used. 
First, the standard rate equations are solved to simulate the primary laser output. 
Second, the secondary laser is modeled with the rate equations that include OIL terms (Eqns. \ref{eq:N_OIL}-\ref{eq:phi_OIL}), using the simulated primary laser output to define $S\inj$ and $\phi\inj$. 
Fig.~\ref{fig:oil} shows the results of one such simulation where a number of secondary pulses are generated during a long, injected primary pulse. 
The secondary pulses during each primary pulse are seen to have a fixed phase, with a constant phase offset between the primary and secondary light. 
Additionally, note the reduction in relaxation oscillations at the start of each pulse compared to the earlier simulations with no injection locking. OIL thus offers a powerful technique for high-speed all-optical phase control of generated pulses.

\subsection{Direct Phase Modulation}
The final laser phenomena we describe is optical phase modulation through current modulation. 
Semiconductor lasers are unique in that their phase is coupled to the carrier density, as shown by the rate equations (Eqn.~\ref{eq:phi}). 
This is because injected carriers change the refractive index of the semiconductor gain medium. 
The refractive index determines the optical path length between the cavity mirrors, and hence the allowed lasing modes. 
Increasing the refractive index leads to an effectively longer cavity, shifting the allowed lasing modes to lower frequencies, and vice versa. 
Controlling the laser carrier density via the applied current therefore gives deterministic control over the laser frequency, and hence phase. 
We can quantify these effects again using the rate equations. A change in frequency is proportional to the time-derivative of the phase:
\begin{equation} \label{eq:chirp_phase}
\Delta \nu(t)=\frac{1}{2 \pi} \frac{d \phi(t)}{d t}
\end{equation}
and Eqs.~\ref{eq:N}--\ref{eq:power} and \ref{eq:chirp_phase} can be combined (neglecting spontaneous emission) to produce a useful expression relating the chirp to the power output \cite{tucker_high-speed_1985}:
\begin{equation} \label{eq:chirp}
\Delta v(t)=\frac{\alpha}{4 \pi}\left(\frac{\mathrm{d}}{\mathrm{d} t}[\ln (P(t))]+ \frac{2\Gamma\varepsilon}{V\eta h\nu} P(t)\right)
\end{equation}
Note that Eqn. \ref{eq:chirp} describes the change in frequency due to changes in the carrier density only. Temperature also affects the refractive index, and hence frequency, of semiconductor lasers. However, changes in temperature occur over slow timescales and the effects on frequency are negligible at the high speed modulations ($>100$~MHz) used for direct phase modulation in quantum communications \cite{tucker_high-speed_1985}. 

The first term in Eqn. \ref{eq:chirp}, proportional to the rate of change of the (natural log) power, represents changes in frequency due to relaxation oscillations of the carrier density after the applied current is suddenly changed. Relaxation oscillations are damped and so this term is called the ``transient'' chirp. If a laser is modulated up and back down, such that the power before and after the modulation is equal, then the phase change induced by the transient chirp is exactly zero. We can therefore focus on the second term, called the ``adiabatic'' chirp, which represents changes in frequency between different steady state values of power. Adiabatic chirp is a consequence of gain compression: at high power outputs non-linear effects reduce the gain. The main effect is spectral hole burning, where spontaneous emission leads to a dip or ``hole" in the inhomogeneously broadened laser gain spectrum at the lasing frequency \cite{huang_gain_1993, petermann_laser_1988, koch_effect_1986}. The carrier density must then increase to recover the same level of gain required for lasing. The carrier density is therefore not exactly clamped above threshold, rather it varies in proportion to the power and the gain compression factor, producing adiabatic chirp. 

\begin{figure}[tbph]
\centering
\includegraphics[width=0.95\columnwidth]{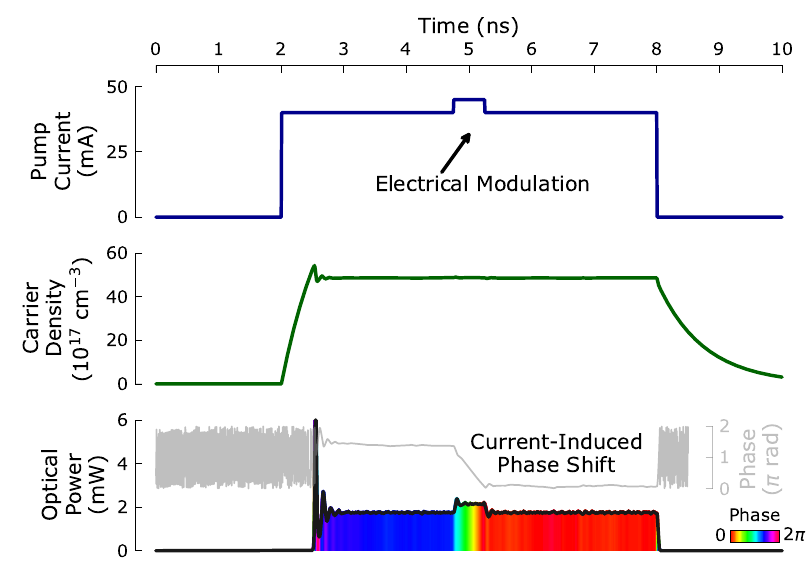}
\caption{Simulated phase modulation of gain-switched pulse by applying perturbation to pump current. By varying the amplitude / duration of the perturbation, the phase shift can be precisely tuned.}
\label{fig:phase_mod}
\end{figure}

We can now express a change in phase as a function of power. 
Considering only the adiabatic chirp, equating Eqns. \ref{eq:chirp_phase} and \ref{eq:chirp}, and integrating gives:
\begin{equation} \label{}
\Delta\phi = \frac{\Gamma\alpha\varepsilon}{V\eta h \nu} \int_t^{t + t_m} P(t) dt
\end{equation}
where $t_m$ is the duration of the modulation. As an approximation we can consider the ideal case where the power is simply proportional to the injected current $ P = \frac{\eta h\nu}{2q} I$ (neglecting the turn-on delay and relaxation oscillations) \cite{tucker_high-speed_1985}. 
For an ideal square pulse modulation, as in Fig. \ref{fig:phase_mod}, the change in phase is therefore:
\begin{equation}\label{eq:phase_modulation}
\Delta\phi = \frac{t_m \Gamma\alpha \varepsilon}{2qV} \Delta I
\end{equation}
For realistic laser parameters (Table \ref{tab:rate_equation_parameters}), and a modulation time of 250~ps, a $\sim$7.4~mA modulation would be required to achieve a $\pi$ phase shift.

Fig.~\ref{fig:phase_mod} shows the rate-equation-simulated dynamics for a semiconductor laser with current-induced phase modulation.
The applied modulation feature at $\sim$5~ns changes the instantaneous laser frequency for a short time period.
This changes the rate of change of phase and thus, Fig.~\ref{fig:phase_mod} shows a quasi-linear phase change during this period (note that phase is plotted relative to a fixed frequency reference).
As a result, the phase of the pulse after modulation can be precisely deterministically controlled by varying the amplitude and duration of the modulation feature.

It is important to note that since the phase is coupled to the laser intensity, direct phase modulation will produce fluctuations in intensity which may be undesirable for applications. 
This issue is circumvented, however, by using an arrangement with the modulated-phase laser as the primary laser, which is injected into a secondary laser, where the secondary emission occurs either side of the modulation feature.
Thus, the pulses that are produced have a clearly defined phase difference and all instantaneous noise associated with the perturbation is rejected.
For example, for time-bin encoding in quantum information, one requires the ability to generate pairs of pulses with precisely defined phase difference between them---this can be achieved by generating two secondary pulses during each primary pulse.
The phase between each pair, however, is random, since between pairs the primary laser is switched off for sufficient duration for cavity photons to deplete, so the subsequent pulse is vacuum-seeded (as described in detail later).

\section{Directly Modulated Laser Sources}\label{sec:direct:modulated:sources}

\subsection{Phase Randomized Pulse Sources for QRNGs and QKD}

The gain switching technique proved of great utility also for quantum communications and it has been used in QKD since its earlier experimental implementations to obtain record transmission rates.

In fact, the most important security requirement of QKD implementations with weak coherent pulses is that light must be strongly attenuated to lower the probability of transmitting more than a photon per pulse. As shown in Fig. \ref{fig:intro:wcp}, for a photon flux of less than 1 photon per pulse the Poissonian distribution is dominated by the 0-photon component, and the receiver has a high probability to measure an empty pulse. The raw photon detection rate is further greatly penalized due to the various losses in the communication channel and instruments. To compensate for the low detection rate, the only solution is to increase the number of pulses transmitted per unit time.

The possibility to generate short pulses is essential also for another reason, still related to the achievable secure key rate.
At the receiver side, single photon detectors can be gated to detect just around a well-defined time interval during which the pulse is expected to arrive.
In this way, the narrower the gating window, the smaller the probability to have false detections due to stray light or dark counts, with a consequent enhancement of the signal to noise ratio.

Most common implementations of QKD transmitters for fiber channels are realized with distributed feedback (DFB) laser diodes with central wavelength at 1300 nm or 1550 nm and the typical pulse length does not exceed 0.5 ns.
Early implementations featured repetition rates in the range of hundreds of MHz.

Most recently, by exploiting the GHz modulation bandwidth of DFB communication lasers, this technique made possible the implementations of high rate BB84 protocols \cite{Dixon.2015,Yuan.2018,Boaron.2018}.
Table \ref{tab:GS_implementations} reports a list of various QKD sources implemented by using the GS lasers.
Notably, GS is applied also with vertical-cavity surface-emitting lasers (VCSEL), which are typically employed for free-space QKD sources.

\begin{table}
\caption{QKD implementations with GS}
\begin{tabular}[htbp]{@{}lllll@{}}
	\hline
	Laser & $\lambda$ (nm) & Pulse width (ps) & Clock Rate (MHz) & Ref. \\
	\hline
	DFB & 1550 & 300 &  & \cite{Hughes.2000} \\
	DFB & 1550 & 80 & 2 & \cite{Gobby.2004} \\
	DFB & 1550 & 400 & & \cite{Yuan.2005}\\
	DFB & 1550 & 100 & 625 & \cite{Tanaka.2008} \\
	DFB & 1550 & 15 & 1000 & \cite{Dixon.2010} \\
	DFB & 850 & 400 & 100 & \cite{Jofre.2010} \\
	VCSEL & 850 & 50 & 1250 & \cite{restelli2009quantum} \\ 
	VCSEL & 850 & 80 & 100 & \cite{WeiChen.2710201302112013b}\\
	VCSEL & 850 & 125 & 100 & \cite{Li.2014b}\\
	DFB & 1550 & 50 & 1000 & \cite{Dixon.2015}\\
	DFB & 1550 & 50 & 625 & \cite{Grunenfelder.2018}\\
	\hline
\end{tabular}
\label{tab:GS_implementations}
\end{table}

Alternative methods that are employed in QKD to generate short pulses are either by using mode-locked lasers or by pulse carving.
Compared to gain switching, these methods present two drawbacks. 
The first is the complexity of the source.
The technique of pulse carving requires indeed the use of a CW laser in connection with external intensity modulators to format the optical field into a train of pulses.
These are either lithium-niobate Mach-Zehnder modulators \cite{Takesue.2005,Honjo.2007,Choi.2010,D.O.Caplan.2018} or electro-absorption modulators \cite{Takesue.2007,Namekata.2011,B.Schrenk.2018}.
The former requires an adequate driving electronics able to supply high voltage short pulses; for the latter the driving signal amplitude can be lower but the price to be paid is a lower extinction ratio.
In order to increase the signal to noise ratio, i.e., the contrast with respect to the background of the photons emitted in CW, it is often necessary to set up a cascade of two intensity modulators, with a consequent doubling of the carving signals, which need to be perfectly synchronized to enable their tandem operation \cite{Lio.,Vagniluca.2020}.

The use of mode-locked lasers would therefore simplify the source \cite{Takesue.2006,Thew.2006} but still this implementation would suffer of the second drawback, i.e., the phase coherence between the train of pulses.
As explained in Sec. \ref{sec:intro}, phase randomization is central for the secure implementation of the WCP BB84 protocol \cite{Gottesman.2004}.
More specifically, security proofs require that each qubit carries a random global phase in order to de-correlate them from possible systems held by Eve \cite{Zhao.2008}.
A phase randomized coherent state is indistinguishable from an incoherent mixture of photon number states and it statistically approximates a single photon source if the amplitude of the coherent field is sufficiently small.

Signals generated either with CW laser carving or mode-locked lasers  without a stage of phase randomization are coherent with each other and therefore cannot meet the experimental assumptions necessary to guarantee the security of the protocols.
The obvious solution to generate a phase randomized state is to use an additional phase modulator to actively encode a random phase on each qubit.
For example this was demonstrated in \cite{Zhao.2007}.
However, the use of external active phase randomization adds an extra layer of complexity to the sources.
First, a source of true randomness is required to select the phases in a uniform and unpredictable way.
The random number generator needs to be fast enough to match the repetition rate of the laser. 
Second, as for the Mach-Zehnder modulators, the lithium-niobate phase modulator, requires fast and high amplitude voltage driving signals.
In particular, the driver has to feature a wide dynamic range and a high resolution in order to faithfully convert the random number into a random voltage for the phase selection.

Compared then to mode-locking or pulse carving, GS is a natural way to bypass the use of the external modulator and greatly simplifies the architecture of BB84 QKD transmitters, as the driving current signal can be as simple as a square wave \cite{Hughes.2000,Yuan.2005,Dixon.2008,Dixon.2010}.
In fact, by recalling Section 2.2, if the amplitude and repetition rate of the driving signal are suitably adjusted, such that the inter-pulse interval is dominated by spontaneous emission, each new pulse starts with a different random phase.
Hence, a GS laser is an ideal source of phase randomized pulses.

\subsection{OIL Sources}

So far, we illustrated how GS represents a versatile technique to obtain short and phase randomized pulses and for this reason it finds a widespread use for BB84 protocols.
However, side effects are also associated with GS, in particular time jitter and frequency chirping.

When the modulation current is injected in the laser diode junction, the emission of the optical pulse is not instantaneous but it occurs after the so-called turn-on delay.
This delay is the time necessary for the carrier population to first reach the threshold value, after which the build-up of the optical pulse begins. 
As the current modulation starts below threshold, the build-up is affected by the randomly fluctuating photon and carrier densities generated by spontaneous recombination.
This translates into time jitter of the optical pulses, as each new pulse is generated with a random relative delay with respect to the current signal (Fig. 11 top). Similarly, the random seed of photons and carriers at the onset of a new optical pulse leads to intensity fluctuations~\cite{Xie.2019}.
In this process of current modulation, also the refractive index of the cavity medium changes and, as explained in Section 2.4, the emission frequency gets chirped with a consequent broadening of the emission spectra.
 
For BB84 protocols, temporal jitter, intensity and spectral diversity have a limited impact because the pulses interfere with themselves.
In fact, although a temporal jitter of 10~ps is of the same order of magnitude as the pulse width, this is anyway lower than the jitter associated to the single photon detectors, at the receiver side.
However, for protocols of more recent introduction, such as the measurement-device-independent one (MDI) based on the interference of pulses generated by two remote laser sources, GS affects the temporal and spectral overlap of the pulses with a dramatic reduction of the interference visibility (see Section 4).

A possible solution to this problem could be pulse carving with active randomization but, as explained earlier, this would greatly increase the complexity of the system.
Recently, it has been discovered that OIL, whose benefits are well known in classical optical communications \cite{Liu.2020b,Lau.2009}, can be elegantly applied to quantum communications.

OIL was initially explored to improve the visibility problem for Hong-Ou-Mandel (HOM) experiment with independent sources \cite{Comandar.2016b}.
Each source consists of two GS lasers: the pulses of one laser optically seeds the other, providing in this way a stable initial density reference of stimulated photons for the pulses of the second laser.
As illustrated in Section 2.4, this condition stabilizes the temporal jitter inherent to spontaneous emission and narrows the spectra of the secondary laser pulses (Fig. 11 bottom).
In addition, the intensity fluctuations of GS pulses are strongly suppressed in the presence of an external seeding field~\cite{Xie.2019}. 

Soon after the HOM experiments, it was realized that OIL could be used to further simplify of architecture of the BB84 transmitter for time bin-encoding \cite{Yuan.2016}. 
In fact, by directly modulating the phase of the primary laser it is possible to encode the qubits in the pulses emitted by the secondary laser, without the need of the interferometer and the phase modulator.
The versatility of this so-called phase-seeding approach is remarkable since it makes possible to implement multiple QKD protocols by simply changing the current modulation format of the primary laser \cite{Roberts.2017,Roberts.2017b,Roberts.2018}, as we will illustrate in next section.

\begin{figure}[h!]
\centering
 \includegraphics[width=\linewidth]{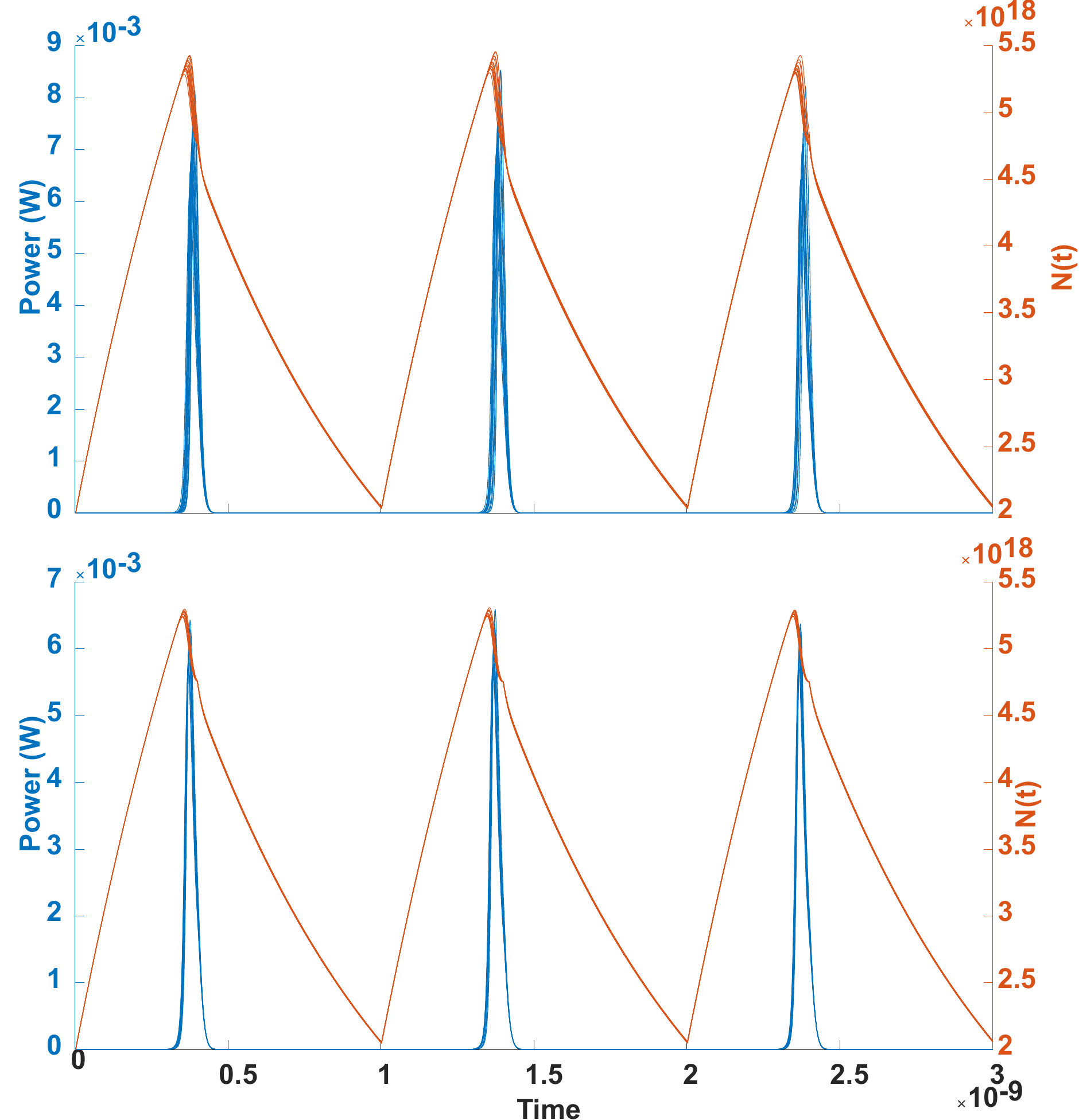}
 \caption{Top: simulation of the time-jitter on the GS switch-on delay. Bottom: Simulation of the time-jitter reduction after the injection of a weak optical field in the secondary laser.}
 \label{fig:dml:gs:timejitter}
\end{figure}

\begin{figure}[h!]
\centering
 \includegraphics[width=\linewidth]{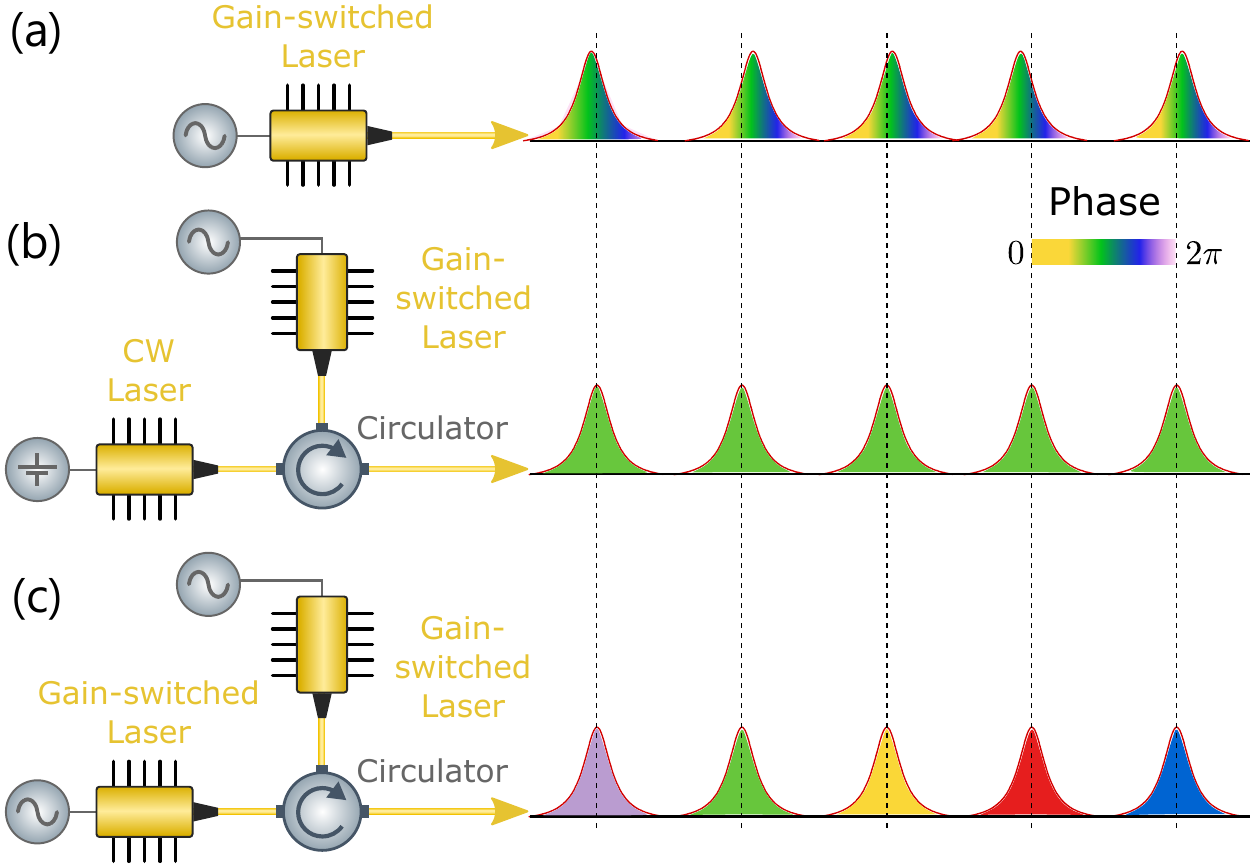}
 \caption{Coherence transfer. Schematics of pulse trains resulting from (a) GS laser:  the short pulses show strong time jitter and do not have a well-defined phase. (b) OIL with a GS secondary laser and a CW primary laser: jitter is suppressed and the pulses all lock to a fixed phase set by the CW seed. (c) Phase randomized OIL of a GS secondary laser by gain switching the primary laser: jitter is suppressed and the short pulses all lock to a different phase set by the phase of the GS primary laser pulses.}
 \label{fig:dml:gs:oil}
\end{figure}

\subsection{Efficient Phase Encoding with Lasers Only}\label{sec:encoding:laser:only}

Figure \ref{fig:dml:gs:oil} summarizes the phenomenology of phase-seeding presented up to now. We have seen how spontaneous emission in gain-switched laser diodes can be exploited to generate short phase-randomized pulses at high repetition rate (Fig.  \ref{fig:dml:gs:oil}~a). We have also seen how it is possible to generate a coherent train of near transform limited short optical pulses with low jitter and low chirp using optical injection locking of the gain-switched pulses with a CW coherent optical seed (Fig.  \ref{fig:dml:gs:oil}~b). Last, we have described how applying a gain-switching modulation to the primary laser itself could then be used to randomize the phase of the pulse train (Fig.~\ref{fig:dml:gs:oil}~c). 
We now move to the description of the quantum state encoding, and see how direct phase modulation can be used to accurately tune the phase of the seed and efficiently encode quantum states for quantum communication applications. 

As discussed in the introduction, encoding a qubit state consists in encoding a polar angle $\theta$ and an azimuthal angle $\varphi$. Writing the qubit state as $c_0|0\rangle+e^{i\varphi}c_1|1\rangle$, the polar angle determines the imbalance in amplitude of the polar states $|0\rangle$, $|1\rangle$ as $c_0 = \cos{(\theta/2)}$ and $c_1 = \sin{(\theta/2)}$, while the azimuthal angle relates to the phase term in the superposition of $|0\rangle$ and $|1\rangle$. The BB84 protocol employs 2 conjugate bases, commonly using 4 equatorial states \cite{Lucamarini.2013}, or 2 equatorial states and 2 polar states \cite{Boaron2018a}. A 6-state version of the BB84 protocol was developed in the late 1990's\cite{H.BechmannPasquinucci.1999}, showing that the use of 3 conjugate bases would further improved the protocol resilience to eavesdropping. The use of 3 conjugate bases, hence 2 polar states and 4 equatorial states, found applications in the early 2010's with the reference-frame independent protocol tailored to tolerate phase/polarization drifts without active compensation \cite{AnthonyLaing.2010,JWabnig.2013,Zhang2014i}. Encoding the polar states $\theta = {0,\pi}$ thus requires completely cancelling either $c_0$ or $c_1$, for example using intensity modulators. Encoding the equatorial states requires acting on the phase of the superposition and this is typically done with phase modulators provided that the two components can be addressed independently.

\subsubsection{Conventional Approaches to Coherent Pulse Encoding}
\begin{figure}[h!]
\centering
 \includegraphics[width=\columnwidth]{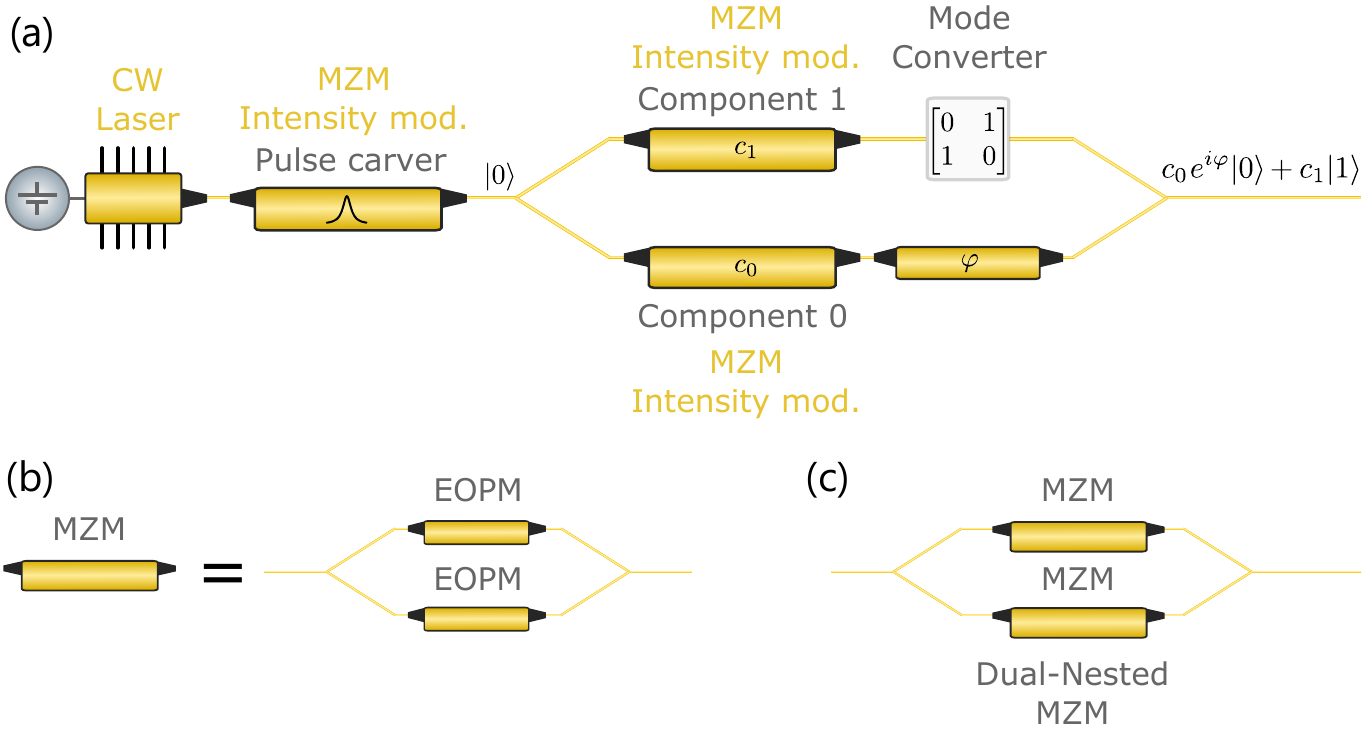}
 \caption{Generic block diagram of a laser-based Bloch sphere encoder. (a) Light from a CW laser is carved into short pulses in a Mach-Zehnder modulator before being encoded in a dual nested MZM, represented here with a phase modulator in one arm and a mode converter in the other. The relative phase between both arms corresponds to the azimuthal angle while the intensity imbalance defines the polar angle in the Bloch sphere (see Sec. \ref{sec:encoding:laser:only}).  (b) A push-pull MZM, consisting of 2 EOPMs in a symmetric Mach-Zehnder interferometer. This configuration is used to reduce the voltage swing needed on a single EOPM. (c) A dual-nested MZM  contains at least 4 EOPMs and hence can be a power hungry component. }
 \label{fig:dml:bloch:sphere:encoder}
\end{figure}

The most generic approach  to encode a coherent pulse in the Bloch sphere is schematically represented in Fig. \ref{fig:dml:bloch:sphere:encoder}: an incoming coherent pulse is split equally into two paths, in which the $|0\rangle$ and $|1\rangle$ components are respectively prepared in parallel by applying the desired intensity and phase modulation in each path before being recombined in a coherent superposition. Depending on the selected encoding, an additional polarization rotation, wavelength shift or time delay is applied to differentiate the $|0\rangle$ and $|1\rangle$ states prior recombination \cite{Bunandar.2018,ChaoxuanMa.2016,PhilipSibson.2017}.

In this review we focus on the states that can be generated with phase modulation only (i.e. the equatorial states, $\theta = \pi/2$) and we leave the discussion of state encoding involving independent intensity modulations of the $|0\rangle$ and $|1\rangle$ states to a future work.

\begin{figure}[h!]
\centering
 \includegraphics[width=0.7\linewidth]{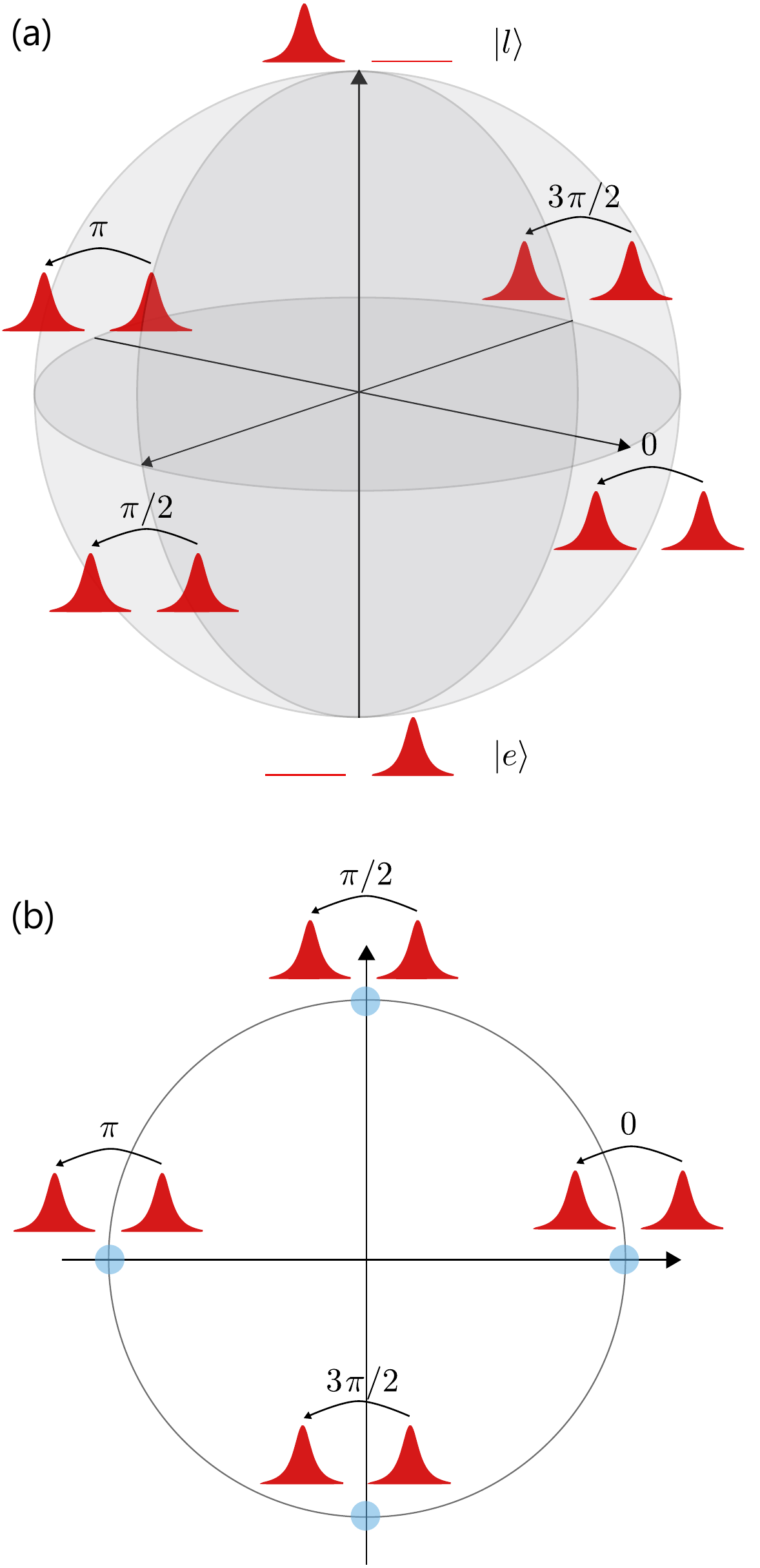} 
 \caption{(a) Time-bin encoding Bloch sphere. The equatorial states are encoded in the differential phase between the early and late pulses. (b) Corresponding representation in the IQ plane. The differential phase $\varphi \in \{0,\pi\}$ for the  $X$-basis states (in-phase) and $\varphi \in \{\pi/2, 3\pi/2\}$ for the $Y$-basis states (quadrature)}.
 \label{fig:dml:time:bin:bloch:sphere}
\end{figure}

Since we are looking at encoding an arbitrary state that is conjugate to the Z basis eigenstates, it is useful to represent states from a polar viewpoint, which corresponds to the projection onto the equatorial plane of the Bloch sphere. This is equivalent to using the In-phase-Quadrature (IQ) plane representation (see Fig. \ref{fig:dml:time:bin:bloch:sphere}b). In this case, we represent the X-basis on the I axis and the Y basis on the Q axis. The radial component of a vector represents the intensity of the state while the angular component represents its phase. The connection with coherent optical communications is worth highlighting \cite{KazuroKikuchi.2016}. Differential quadrature phase shift keying (DQPSK), i.e. the encoding of information in the differential phase of optical pulses is a widely used modulation format in digital communications. A typical IQ-modulator encodes separately the intensities of the I and Q quadratures of the signal similarly to what is described above. Incoming pulses from a coherent pulse source are split into two parallel paths of a symmetric MZI: path 1 applies a $\pi/2$ phase shift to convert from I to Q quadrature. In each path, intensity modulation is used to encode the final components before recombining the pulses.

Here we show how a differential phase can be encoded between successive pulses without the need for splitting the pulse train into independent paths and applying independent modulations. Rather, we present a method that exploits the most of the laser physics described in Section \ref{sec:theory} to generate a phase-encoded stream of pulses with arbitrary differential phase shifts while preserving spectral and intensity properties compatible with QKD requirements. The pulse stream hence generated can readily be used for time-bin encoded QKD protocols. The same stream of pulses could also be used for polarization encoded QKD by combining it with schemes for time-bin to polarization conversion of qubits \cite{ConnorKupchak.2017,Vasconcelos2020,Anderson2020}.

We follow the conventional terminology and refer to each pair of consecutive pulses as one communication symbol. Depending on the phase encoding technique, a single symbol can encode for multiple logical bits. The information is decoded in a demodulator, which typically consists of a pair of delay-line interferometers, used to extract the signal amplitude along each quadrature by interfering the early pulse (reference pulse) with the late pulse (signal pulse) in each symbol \cite{KazuroKikuchi.2016}.

\subsubsection{Tunable Coherence Transfer using OIL and Direct Modulation}
\begin{figure}[h!]
\centering
 \includegraphics[width=\columnwidth]{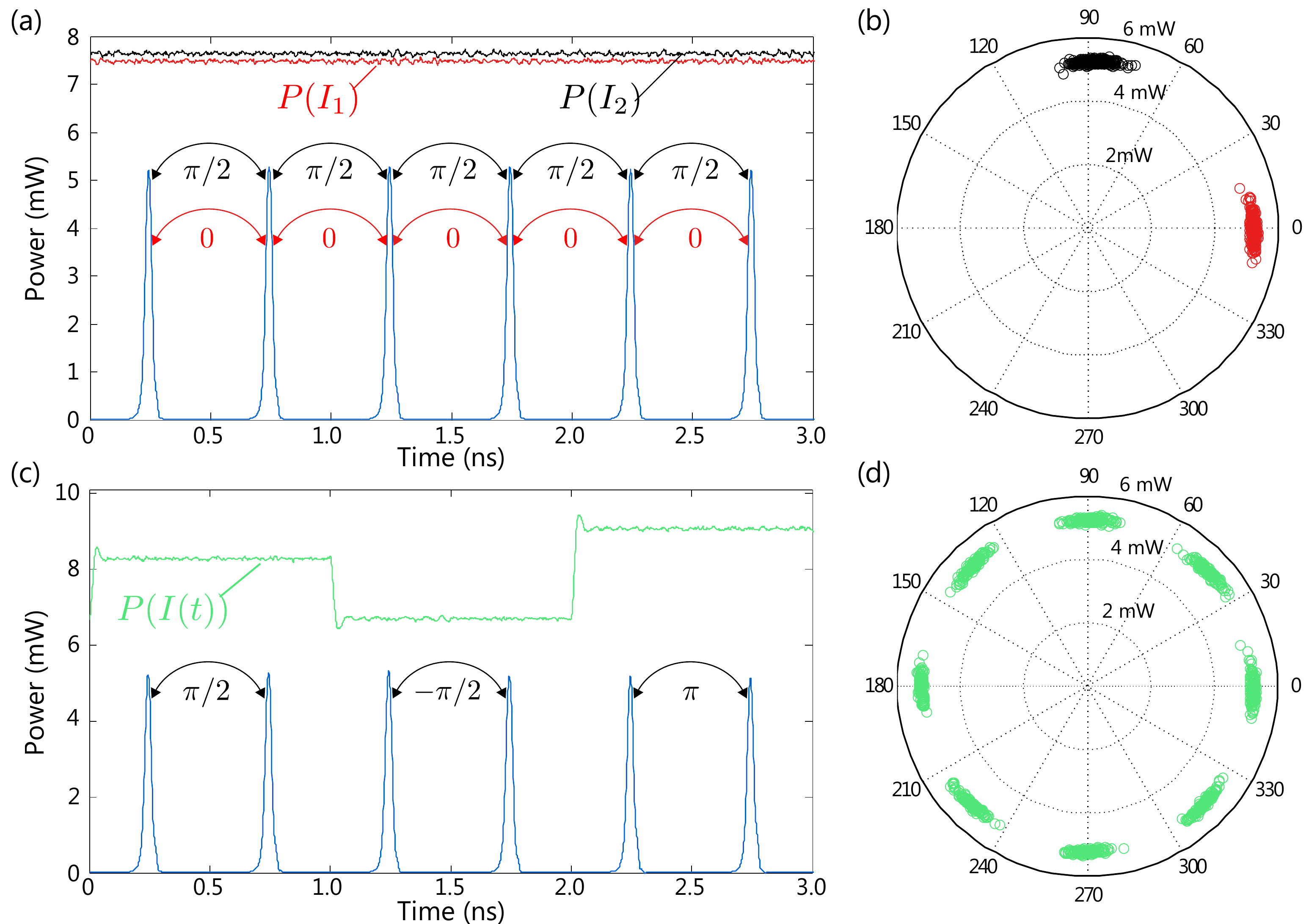}
 \caption{Phase-seeding with CW primary laser and GS secondary laser. (a): Waveforms showing the time-dependent optical power of the CW primary laser, simulated for 2 different bias currents, $I_1$ (red) and $I_2$ (black), and of the short pulses generated by gain-switching the secondary laser (blue). The induced differential phases induced between successive pulses via OIL are shown with the double-sided arrows, for both cases, $I_1$ and $I_2$. (b) Simulated constellation plots of the secondary laser pulse train as decoded in a demodulator. The two distinct spots correspond to the cases of pulses seeded by a primary lasers biased with $I_1$ (red) and $I_2$ (black), respectively. (c) Simulated waveforms showing the optical power of a primary laser driven with direct current modulations $I(t)$ (green) and a gain-switched secondary laser (blue). The encoded differential phases are shown with double-sided arrows. (d) Constellation plot of the differential phase encoded in the secondary laser pulses (8 phases were encoded in this case, corresponding to an 8-DPSK signal).}
 \label{fig:seeding:cases}
\end{figure}

We start with the case of a train of short coherent pulses generated by OIL of a GS laser with a continuous wave seed. As described in Section~\ref{sec:OIL}, if the stimulated emission of the injected light is sufficient to overcome spontaneous emission, the phase of the GS pulses of the secondary laser, otherwise random, locks deterministically to the phase of the primary laser's field present in the secondary cavity at the time of the pulse generation. 
As shown in Eq. \ref{eq:chirp_phase}, the differential phase $\Delta \phi$ between two successive GS pulses is given by the phase evolution of the primary laser's field during a time $\Delta t = T$, the period of the secondary laser. Hence, $\Delta \phi = \nu_P T$, with $\nu_P$ the optical frequency of the primary laser. This is illustrated in Fig. \ref{fig:seeding:cases}a where we simulate OIL with a CW primary laser seeding a secondary laser gain-switched at a 2 GHz repetition rate. 
Measuring the differential phase between consecutive pulses in a demodulator would result in a single phase. This phase depends on the detuning between the primary and secondary lasers and on the time difference between successive pulses. 
Running the simulation with 2 different bias currents of the CW primary laser yields 2 different differential phases, both represented in the IQ plot of Figure~\ref{fig:seeding:cases}b.

\subsubsection{Phase-Seeding and Phase Randomization}

\paragraph{Phase-Seeding}
We now want to encode information in the differential phases of the pulses. Following the previous example, a straightforward way is to use direct modulation of the primary laser by modulating the drive current in each symbol (duration covering 2 pulses of the secondary). The principle is shown in Fig.~\ref{fig:seeding:cases}c and d, in the case of 8 encoded differential phases.
While this approach may be suitable for conventional optical communications, it introduces side-channels for QKD. As described in Sec. \ref{sec:theory}, the frequency of laser diodes is coupled to the emitted power (see Eq. \ref{eq:chirp}). The intensity of the secondary laser pulses also depends on the injected power \cite{huang_laser-seeding_2019}. If each symbol is seeded with light of a different intensity then the phase information becomes correlated with the frequency and intensity, which may constitute a side-channel.

A viable strategy is to modulate the primary laser with a current modulation of duration (duty cycle) much shorter than the symbol duration, and to synchronize this short modulation with the center of the secondary pulse pair.
This guarantees that the pulses in the secondary laser are seeded by a field of constant amplitude and optical frequency and hence no side-channel is introduced by the modulation. By precisely calibrating the amplitude and the duration of the short direct current modulation, it is possible to seamlessly encode the train of secondary laser pulses with deterministic differential phases. This is simulated in Figure 16a and b, where a pseudo-random 8-level short modulation is used to encode a differential phase in each symbol.

\paragraph{Pulsed Phase-Seeding}
The final example introduces phase randomization of the reference pulse of every consecutive pulse pair. Compared to the previous case where a phase is encoded in each symbol, here signal symbols (encoding logical bits) and random symbols (not used to encode information) alternate periodically. One could consider encoding a random phase shift in every other symbol, however for the randomization to be efficient, this would require encoding a large number of different phase states, which would be costly in terms of driving electronics. Instead, one can again make use of the intrinsic phase randomization incurring to the gain-switch process. 

\begin{figure}[h!]
\centering
 \includegraphics[width=1\columnwidth]{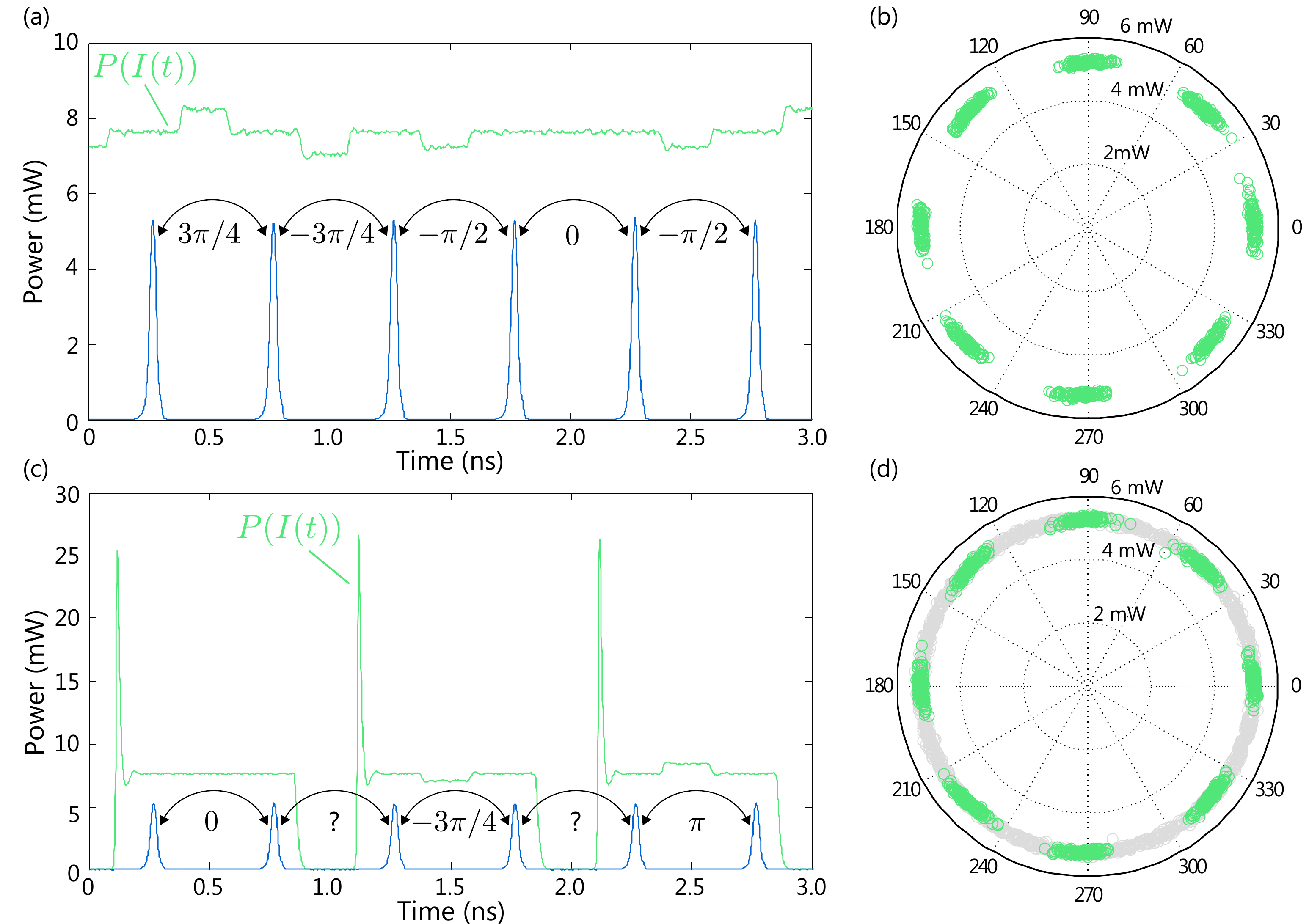}
 \caption{Phase encoding with short direct modulations of the seed laser. (a, b) Waveforms and constellation of a phase-encoded pulse train using short 8-level direct modulation of the primary laser with 20$\%$ duty-cycle current modulations. Blue: pulse train of the secondary laser. Double-sided arrows: encoded differential phases. (c) Pulsed phase-seeding, combining gain-switching with short direct current modulations of the primary laser. Gain-switching the primary laser between each symbol randomizes the global phase, i.e.  the phase between consecutive pairs of pulses is random (question marks).  (d) Corresponding constellation of differential phases as decoded in a demodulator. Green: constellation obtained from the  phase encoded pairs of pulses. Gray: Accounting for the global phase, the constellation spans uniformly all angles. It is no longer possible to distinguish the encoded states using a fixed phase reference. }
 \label{fig:seeding:constellations}
\end{figure}

The idea, simulated in Fig. \ref{fig:seeding:constellations}~c, consists in generating a train of gain switched pulses from the primary laser at half the secondary laser's repetition rate and with a long duty cycle such that a single pulse from the primary laser is able to seed 2 pulses from the secondary laser, while being phase-uncorrelated from its neighboring pulses. A short direct phase modulation is applied within each primary laser pulse to encode a short local phase shift between the regions of the primary pulse that seed the phase of the pair of secondary pulses, as described in Fig. \ref{fig:phase_mod}. 
An important trade-off is that the primary laser's duty-cycle should be \emph{long enough} such that relaxation oscillations are negligible in the seeding region, and \emph{short enough} for the carrier number to deplete sufficiently between two pulses such that the amount of stimulated emission contributing to the generation of a new pulse is negligible (see Sec. \ref{sec:OIL}). In addition, the phase-tuning modulation should be \emph{short enough} to avoid temporal overlap with the secondary pulses, but \emph{long enough} to keep the modulation amplitude low. For instance, in Fig. \ref{fig:seeding:constellations}~c, the primary laser repetition rate is half that of the secondary laser with a 0.8 duty cycle, and the phase-tuning modulation has a duty-cycle of 0.2 and a half-wave current $I_\pi = 10$~mA, which is consistent with the 9.25~mA expected from Eq. \ref{eq:phase_modulation} since the carrier dynamics would not directly follow the square  current modulation but rather involve some transient.

\section{Applications and Experimental Demonstrations}\label{sec:applications}

\subsection{Quantum Random Number Generation}
The same feature that is exploited to remove the burden of the active phase randomization stage in the transmitter, i.e., the inherent random phase of GS pulses, can be used to implement sources of true randomness that are needed for the encoding and decoding of the qubits.

\begin{figure}[h!]
\centering
 \includegraphics[width=\linewidth]{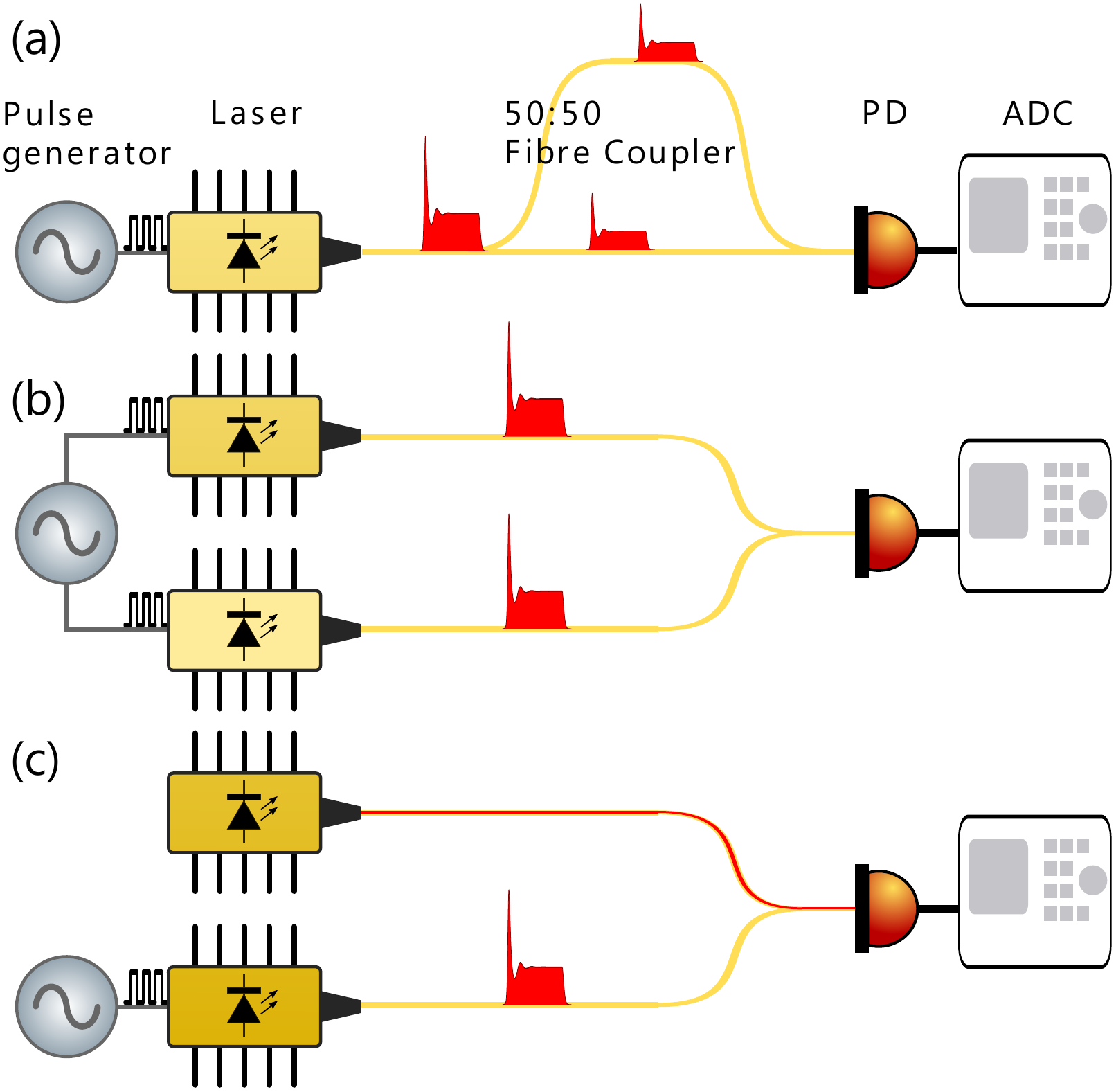}
 \caption{Basic setups to implement quantum random number generators by exploiting gain switched lasers. a) A laser is modulated to obtain phase randomized pulses that are then split and retarded with a delay matching the modulation repetition period by an asymmetric interferometer. The output pulse features a random amplitude which is converted into a random current signal by a photodiode (PD). This signal is then converted into a digital value by an analogue-to-digital converter (ADC). The interferometer can be replaced by the direct interference of  fields generated from independent lasers, where either both  lasers are gain switched (b) or one is operated in CW (c). }
 \label{fig:gs:QRNGs}
\end{figure}

Quantum random number generators (QRNG) can then be easily implemented by exploiting the simple fact that the product of the interference of two pulses with random phases is a third pulse with a random amplitude.
The random amplitude of the optical field can then be converted into a random current signal by means of a photodiode (PD) and then random numbers are then generated by sampling the electrical signal with an analogue to digital converter (ADC). 
In Fig. \ref{fig:gs:QRNGs} three possible arrangements for the generation scheme are summarized: all the schemes interfere two optical fields, the difference being the way they are generated.
The simplest and most common arrangement, Fig. \ref{fig:gs:QRNGs}-a, features a single gain switched laser and interference is obtained by means of an asymmetric interferometer with a delay matching the repetition period of the laser modulation signal.
This scheme was originally introduced in \cite{Abellan.2014} and \cite{Yuan.2014}.
The random amplitude optical field can be generated by direct interference of two independent lasers, either both gain switched  \cite{Sun.2017}, Fig. \ref{fig:gs:QRNGs}-b  or with one of the two operated in CW Fig. \ref{fig:gs:QRNGs}-c  \cite{Abellan.2016,Sun.2017}. 
The two-laser design adds a degree of complexity since the emission wavelengths have to be matched in order to obtain high-visibility interference. 
However, this scheme turns out to be more advantageous than the single laser one, when realized with integrated photonic chips
\cite{Abellan.2016,ThomasRoger.2019}.
In fact, the integrated delay line introduces high losses with respect to the short arm of the interferometer creating in this way an unbalance between the two arms, which needs to be compensated either by adding additional losses in the short arm \cite{MuhammadImran.2021} or by splitting light with uneven ratio at the input of the interferometer \cite{Rude.2018}.
As explained in Section 3.1, DFB lasers can be modulated at very high rates such that, if the laser is matched with a high bandwidth PD and a fast ADC, bit generation rates in the range of hundreds of Mbit/s and tens of Gbit/s can be achieved \cite{Abellan.2015,Marangon.2018}.

\subsection{Multi-Level Optical Communications}
\begin{figure}[h!]
\centering
 \includegraphics[width=\linewidth]{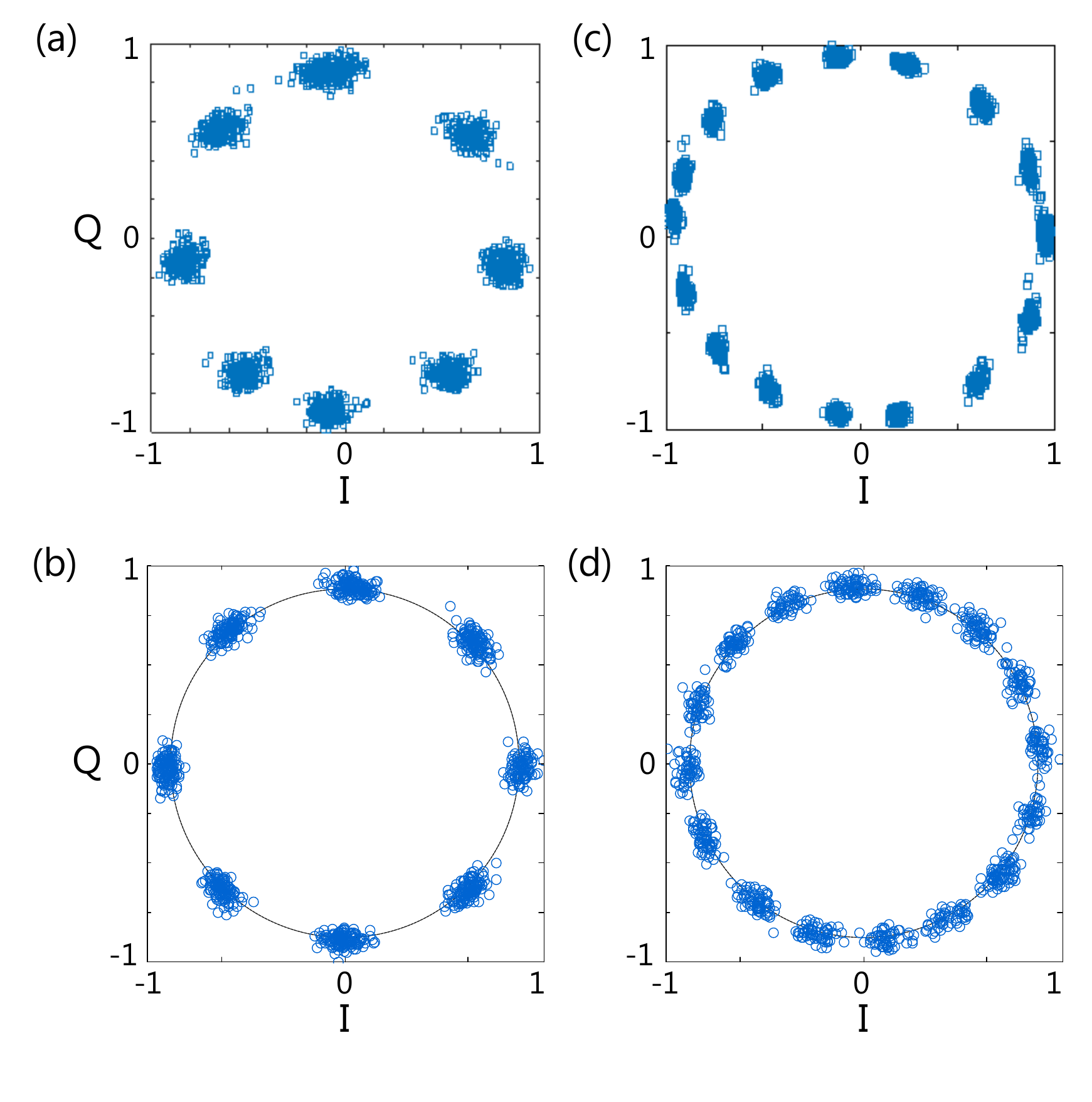}
 \caption{Multilevel (RZ) M-DPSK signalling using a direct phase modulation source. (a, b) $M = 8$, 3 bit per symbol, experiment and simulation, respectively. (c, d) $M = 16$, 4 bit per symbol, experiment and simulation, respectively. Top row: experiment. Bottom row: simulation.}
 \label{fig:seeding:constellation:experiment}
\end{figure}

Introduced in the early 2000s \cite{R.A.Griffin.2002,P.J.Winzer.2006}, multi-level (or \emph{advanced}), optical modulation formats have become essential for high-capacity fiber communications. The alphabet size (i.e. the number $M$ of states that a pulse can be encoded into) is increased to emit more than 1 logical bit per symbol. The  number of logical bits per symbol increases with  the alphabet-size as $\log_2(M)$. Return-to-zero (RZ) signalling using light pulses is also convenient to reduce inter-symbol interferences \cite{A.H.Gnauck.2005}.
Figure \ref{fig:seeding:constellation:experiment} presents experimental realizations of the RZ-8DPSK and RZ-16DPSK modulation formats, encoding for 3 bits and 4 bits of information per symbol, respectively. In these experiments, the pulses were encoded using a direct-phase modulated source with a symbol rate of 2 GHz, yielding data bit rates of 6 Gb/s and 8 Gb/s, respectively. In this experiment, DFB lasers with a 15 GHz analogue bandwidth were used.
With DFB lasers now reaching bandwidths of several 10s GHz, this approach provides an attractive way of implementing advanced modulation format at high bit rate in a cost-effective and electronics-efficient way. The simulated constellations shown in Fig.~\ref{fig:seeding:constellation:experiment} are in very good qualitative agreement with the experiments. The laser parameters are the ones of Table \ref{tab:rate_equation_parameters}.

\paragraph{Imperfect Phase Encoding}
As can be seen from Fig.~\ref{fig:seeding:constellation:experiment}, due to noise in the laser and the driving electronics, the encoded phases form a narrow distribution around the ideal value.
Imperfect phase encoding leads to errors when measured, increasing the bit error rate (BER). This also limits the number of distinct phase values that can be decoded reliably by the measurement device. Parameters such as the laser linewidth and the spontaneous emission noise play an important role in the achievable sensitivity at the detector \cite{M.Seimetz.2008}. To further increase of the sensitivity, the distance between the different states of the constellation can be increased by combining the M-DPSK modulation with amplitude modulation in quadrature-amplitude modulation (QAM) \cite{KazuroKikuchi.2016}. This would require appending an intensity modulator to the phase seeded transmitter.

\begin{figure}[h!]
\centering
 \includegraphics[width=\linewidth]{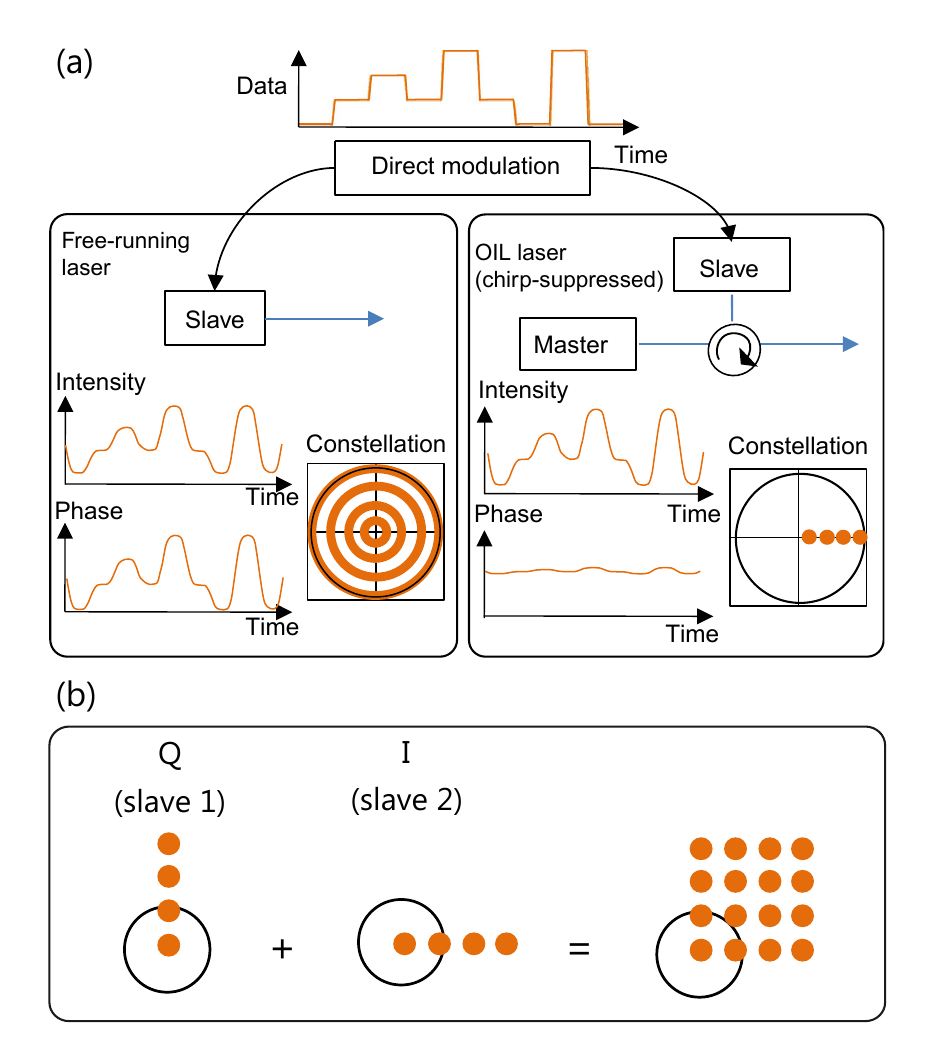}
 \caption{Direct laser modulation for multi-level optical communications. a) In this approach the phase of the secondary laser (slave) is locked to the phase of the primary laser (master) and intensity modulations of the secondary laser creates a fixed phase constellation. b) By combining the signal from two secondary lasers with an appropriate phase shift, the authors were able to encode orthogonal quadratures and generate a 16-quadrature amplitude modulation signal. Adapted from Ref.~\cite{Liu2014g} with permission. 2014, Springer Nature.
}
 \label{fig:applications:classical:comm}
\end{figure}

\subsection{Point-to-Point QKD}
A great variety of QKD protocols has been devised since the original BB84 protocol. Here we review how a direct phase modulated light source can be used as a transmitter for some of the most common protocols.
Protocols such as BB84, B92,SARG04, and 6-state protocol are notable examples of discrete variable (DV) protocols and are described in great details in several reviews \cite{Scarani.2009,S.Pirandola.2020}. DV protocols encode information in the eigenmodes of selected observables. They are implemented with single photons or WCPs and therefore require single photon detection. As mentioned in Section \ref{sec:intro} these protocols generally require phase randomization to guarantee security over large distances \cite{Lo.2007}. 

Another class of protocols, called the distributed phase reference (DPR) protocols, is also based on single photon detection but does not require phase randomization. Rather, DPR protocols exploit the photon statistics in a stream of weak coherent pulses. The information is encoded in the intensity (coherent one way protocol  \cite{Stucki2005}) or in the relative phase (differential phase shift protocol \cite{Inoue.2003}) between consecutive pulses of a coherent pulse train. The occurrence of single photon detection events is unpredictable due to photons being effectively delocalized over several pulses. The security comes from the fact that an attacker trying to intercept and resend part of the signal would unavoidably alter the signal's coherence. Figure \ref{fig:applications:qkd:cases}(a-c) presents the output of QKD transmitters for the COW, DPS and BB84 protocols, respectively. 

The requirements to realize a QKD transmitter are presented in the generic block diagram of a QKD transmitter shown in Fig. \ref{fig:applications:qkd:cases}d. Three important functions are needed: a coherent light source, a pulse generator and a quantum encoder. In addition, the variables used to encode the photon states must be completely unpredictable, which is typically achieved using quantum random number generators.

Owing to its flexibility for encoding complex phase states, the phase-seeded transmitter is an attractive light source for QKD. Once the key information encoded in the photon states and the pulses attenuated to a mean photon number $\mu < 1$, the same transmitter can be used to implement different protocols, without the need for setup reconfiguration.  

Figure \ref{fig:applications:driving:cases} shows how the COW, DPS and BB84 states are generated by simply changing the driving conditions of the phase-seeded transmitter, using the physics that we described in Section \ref{sec:direct:modulated:sources}.

\subsection{Distributed Phase Reference Protocols}
Alice and Bob exchange a stream of coherent pulses following a Poissonian distribution with mean photon number $\mu < 1$.  The security of the protocols arises from the fact that Eve cannot precisely measure the phase and the photon number at the same time. Therefore Eve cannot predict in advance the occurrence of detection events at Bob. In particular, since the phase and photon number coherence is delocalized over the whole pulse train, if Eve attempts to measure the phase (the photon number) of successive pulses, she necessarily loses the photon number (the phase) information and therefore introduces measurable errors in the communication. These protocols conveniently offer high key rates however a complete proof has not yet been established.

\subsubsection{Coherent One Way Protocol}
In the coherent one way (COW) protocol the logical bits are encoded in the $Z$-basis, and a single state of a conjugate basis, say the $X$-basis, is used to monitor the presence of an eavesdropper. Alice prepares $|0_Z\rangle$,  $|1_Z\rangle$ and $|0_X\rangle$ Bob randomly chooses to measure the symbols either in the $Z$-basis to obtain key bits, or in the $X$-basis to measure the phase coherence of the signal. At the end of the transmission, Bob reveals which symbol were measured in which basis and reconciles the key information with Alice. The QBER in $X$ is used to detect the eavesdropper \cite{Stucki2005}.

If Eve attempts to measure the photon number to predict the $Z$-basis outcome, then she destroys the phase coherence and degrades $X$-basis measurements. Conversely, if Eve attempts to measure the phase coherence between successive pulses, she loses the photon number information. 

\begin{figure}[h!]
\centering
 \includegraphics[width=\linewidth]{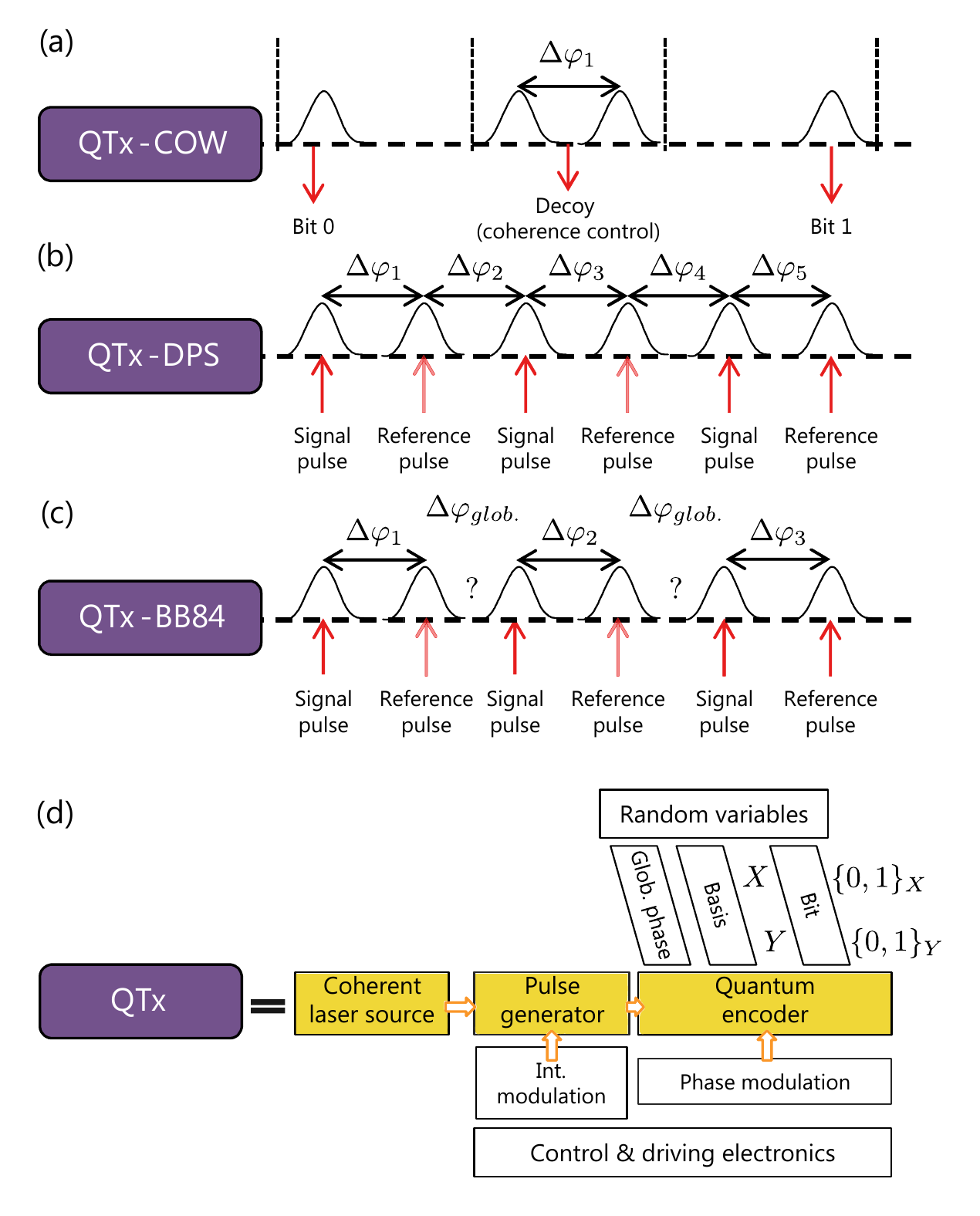}
 \caption{Signals from QKD transmitter for different protocols. (a) COW, (b) DPS, (c) BB84, (d) Generic QKD transmitter.}
 \label{fig:applications:qkd:cases}
\end{figure}

\begin{figure}[h!]
\centering
 \includegraphics[width=\linewidth]{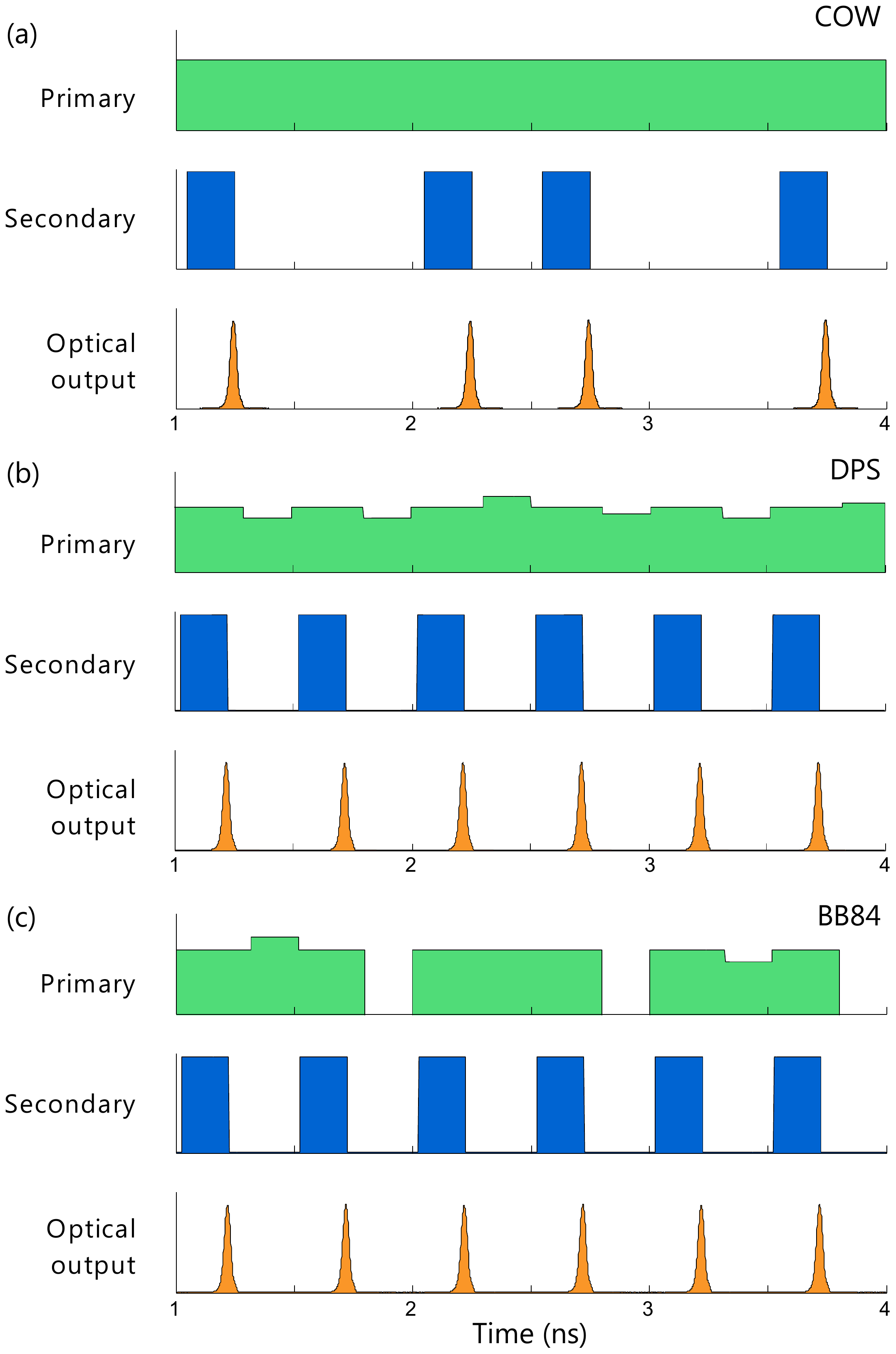}
 \caption{Driving signals to operate the phase-seeded QKD transmitter for 3 different protocols. (a) COW, (b) DPS, (c) BB84.}
 \label{fig:applications:driving:cases}
\end{figure}

\subsubsection{COW Implementation with a Phase-Seeded Transmitter}
The COW transmitter needs to output a stream of fully coherent pulses with a single phase relation $\Delta \varphi_1$ between consecutive pulses. However, some pulses of the pulse train are empty while some others are prepared with a mean photon number $\mu<1$ (see Fig. \ref{fig:applications:qkd:cases}~a). The sequence of $\mu$-pulse – empty pulses encodes for the logical bit 0 and the sequence empty-pulse - $\mu$-pulse encodes for the logical bit 1. Two consecutive $\mu$-pulses encode a decoy pulse, which if measured in the $X$-basis provides a measure of the pulse train coherence.

Only the coherent light source and intensity modulation functions are required. The transmitter must be capable of emitting short pulses with low time jitter (important for the $Z$-basis measurement) and high phase coherence (important for the $X$-basis measurement). A straightforward realization of the COW transmitter would only require a CW laser and a high-speed pulse carver. In Fig. \ref{fig:applications:driving:cases}~a we show how this is implemented with the phase-seeded source: the primary laser is driven in CW and injected into the gain-switched secondary laser, driven with binary (ON-OFF) data. The main advantages compared to the pulse carver approach are a greater extinction ratio and shorter pulses. The coherent pulse train is then attenuated to the desired average photon flux $\mu$, e.g. using a fixed optical attenuator, before being sent to the receiver. An experimental realization of the COW protocol with a phase-seeded transmitter can be found in \cite{Roberts.2017}, where a secure key rate could be generated up to 30 dB attenuation, corresponding to a distance of 150~km of standard single mode fiber (see Fig. \ref{fig:applications:discrete:optics}~a).

\subsubsection{Differential Phase Shift Protocol}
In the differential phase shift (DPS) protocol, Alice encodes the logical bits in the $X$-basis as $|0_X\rangle$ and $|1_X\rangle$. All the pulses contain an average photon number $\mu<1$. Bob only measures the $X$-basis, however, due to the Poissonian statistics, he will only detect signals at random times. At the end of the transmission Bob reveals at which time slots he measured a signal and proceeds to the reconciliation with Alice \cite{Inoue.2003}. 

The security arises from the fact that Eve cannot predict when a detection event (click) will happen at Bob’s receiver. Nor can she force a detection event in a single time slot: if Eve attempts an intercept and resend attack, she would need to prepare two pulses with a photon number sufficient to ‘force’ a click at the desired time slot, but she would remain unable to prevent the statistical occurrence of clicks in the neighboring time slots, thus increasing the QBER. Conversely, if Eve were able to measure the photon number in each pulse, she would then project each pulse onto a Fock state and thereby destroy the phase coherence of the pulse train. Moreover she still would not be able to predict the clicks at Bob.\cite{Inoue.2003}

\subsubsection{DPS Implementation with a Phase-Seeded Transmitter}
The DPS transmitter needs to output a stream of fully coherent pulses and control the differential phase $\Delta \varphi$ between consecutive pulses. All pulses of the pulse train are prepared with the same mean photon number $\mu<1$. A differential phase $\Delta \varphi=0$ encodes for the logical bit 0 while $\Delta \varphi=\pi$  encodes for the logical bit 1. Because $\mu<1$, pairs of consecutive pulses would only  contribute to detection events with a finite probability.

Here again, the target is first to generate a train or short pulses with high phase coherence, low chirp and low time jitter. In addition, a quantum encoder function is required to prepare the different states. Realizations using a laser, pulse carver and IQ modulator similar to the DPSK modulator have been demonstrated in various works. 
With the phase-seeded source, a pulse train of short pulses can readily be encoded in phase using OIL of a phase encoded seed prepared through short direct current modulations of a CW primary laser, as described in Fig. \ref{fig:seeding:constellations}a. For the DPS protocol only 2 phase states $\{0, pi\}$ are needed, and it is therefore sufficient to use only 2 levels of short amplitude modulation $I_{\textnormal{mod}}=0,\ I_\pi$. An experimental demonstration of the DPS protocol was given in \cite{Yuan.2016}. The states were encoded with a low half-wave voltage $V_\pi = 0.35$~V corresponding to modulation current as low as $I_\pi = 7$~mA considering a 50~$\Omega$ load.

\subsubsection{Advanced DPS Protocols}
Interestingly, the DPS protocol can be generalized to 4 phase states in the differential quadrature phase shift (DQPS) protocol, thus corresponding to the DQPSK modulation format of classical communications \cite{KyoInoue.2009}. As we discuss below, phase randomization further enhances the security of the protocol.  

The DPS protocol is also related to continuous variable QKD protocols (CV-QKD) with discrete modulation, where coherent detection is used to measure the states. An overlap between the constellation points introduces the state ambiguity used to establish the protocol security \cite{ShouvikGhorai.2019}. 

\begin{figure}[h!]
	\centering
	\includegraphics[width=\columnwidth]{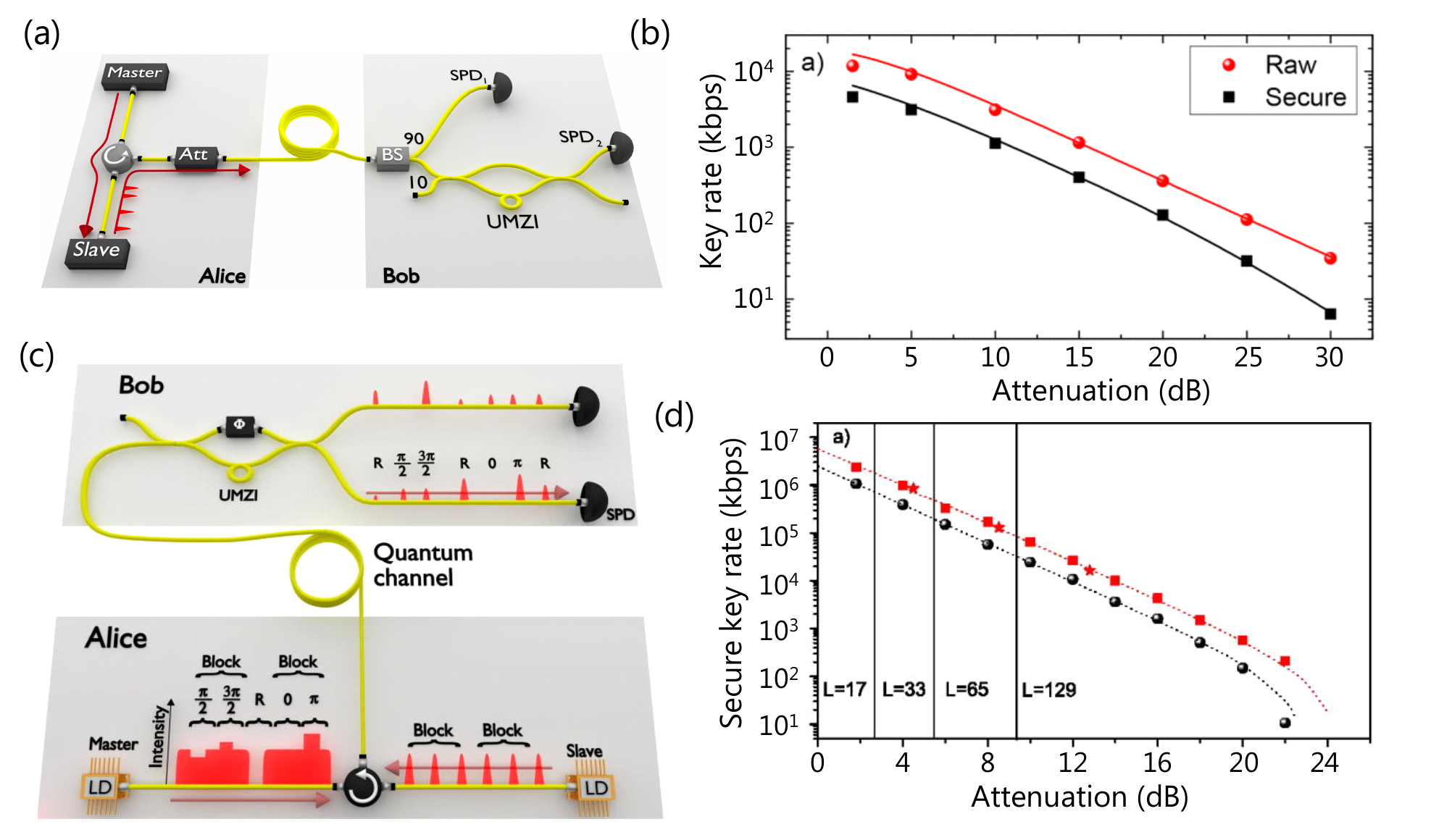}
	\caption{QKD experiments with a phase-seeded source. (a-b) Experimental realization of the COW protocol. (\emph{Adapted from Ref.~\cite{Roberts.2017}}). (c-d) DQPS protocol (red) and performance comparison with the BB84 protocol (black). (\emph{Adapted from Ref. \cite{GeorgeL.Roberts.2017}}).}
	\label{fig:applications:discrete:optics}
\end{figure}

\subsection{Phase Randomized BB84 protocol}
In contrast with the DPR protocols, DV protocols often require the global phase to be randomized regularly. The main motivation, as discussed in Section \ref{sec:intro}, is to map the density matrix of the emitted weak coherent states to that of a mixture of Fock states with coefficients taken from a Poisson distribution of mean $\mu<1$.

As shown in Fig. \ref{fig:intro:phase:rand}, operating the BB84 protocol beyond the 10~km range is only achievable in the presence of phase randomization. Even longer distance QKD was enabled by the decoy-state protocols, which combine phase randomization and  weak coherent pulse intensities selected at random out of a well-defined set in order to mimic the preparation of a mixture of photon number states with variable Poissonian coefficients. This approach provides a way to estimate the expected yield of single photon pulses and detect a potential photon number splitting attack by an eavesdropper \cite{Hwang2003,HoiKwongLo.2005}.

Thanks to the phase randomization and the mapping to photon number eigenstates, the statistical contributions from single photon  pulses and multi-photon pulses can be estimated in the security analysis. In addition, since 2 bases are used, the sifting and verification procedure leaves Alice and Bob with two bit error rates, $\delta_X$ and $\delta_Y$, one contributing to the bit error correction and the other to the `phase error' correction \cite{A.R.Calderbank.1996,ASteane.1996}. 

A simplified argument is to consider the \emph{intercept-and-resend} attack (see for example Ref. \cite{S.Pirandola.2020}), where Eve makes her measurements in one basis, say the $X$ basis and re-emits pulses prepared in the states she measured. 
Assuming Alice encodes $X$-basis states, Eve's action might alter the phase of the states, e.g. by changing $|1_X\rangle$ into $-|1_X\rangle$, however, this would not cause bit-flip errors. These errors, called phase errors, correspond to Eve gaining information on the key without causing bit flips. If on the other hand Alice encodes states in the $Y$-basis, then Eve's action results in bit-flip errors with a 50 $\%$ probability because it would project the incoming $Y$ basis states onto an $X$ basis state prior Bob's measurement. Since Eve cannot predict the encoding/decoding bases of Alice and Bob, it is assumed that Eve attacks all pulses uniformly independent of their actual basis, and therefore the bit error rate in one basis serves as a witness for the phase error rate in the other basis. In our example, $\delta_Y$ can be used to quantify how much Eve has attacked and gained information about the key~\cite{PeterW.Shor.2000,Koashi.2005,Gottesman.2004}. 

Note that $X$ and $Y$ have interchangeable roles, although optimized versions of the protocol use asymmetric basis selections, with one basis used to transmit the key and the other used for phase-error measurements.
After sifting, a first error correction is applied to correct the bit flips and a second error correction (privacy amplification) is applied to correct for the phase errors and completely remove Eve's knowledge about key bits.

\subsubsection{BB84 Implementation with a Phase-Seeded Transmitter}
Encoding states for the phase-encoded BB84 protocol requires encoding 4 phases and periodically randomizing global phase (see Section \ref{sec:intro}). Hence, the transmitter emits pairs of pulses such that there is no deterministic phase relation between consecutive pairs of pulses. Each pair is prepared with a mean photon number $\mu < 1$, typically of the order of 0.5 photon per pulse (see corresponding Poissonian distribution in Fig. \ref{fig:intro:wcp}). Within a given pulse pair the differential phase encodes 2 bases and 2 logical bits, i.e. 4 states $|0_X\rangle$, $|1_X\rangle$, $|0_Y\rangle$ and $|1_Y\rangle$ corresponding to the differential phases $\Delta\varphi = \{0, \pi, \pi/2, 3\pi/2\}$, respectively. 

Using a phase-seeded transmitter, a phase-encoded pulse is prepared with 4-level short modulations to encode the seed. In addition every other clock cycle, the seed is phase randomized by gain-switching the primary laser. The seed is then injected into the secondary laser, in turn gain-switched with a regular square signal as shown in Fig. \ref{fig:applications:driving:cases}c. The decoy-state protocol mentioned above can further be realized by appending an intensity modulator at the output of the secondary laser.

\subsubsection{Phase-Randomized DQPS Protocols}
A hybrid protocol between the DQPS and the BB84 protocol, including phase randomization, has  also been demonstrated with a phase-seeded QKD transmitter, bridging the security gap with the BB84 protocol. In this protocol, phase-coherent signal pulses are emitted per blocks of length $L \geq 2$, with $L=2$ being equivalent to the BB84 protocol \cite{ShunKawakami.2016}. To realize this, the primary laser generates a gain-switched pulse long enough to seed $L$ secondary laser pulses. Each block of $L$ pulses contains an average of $\mu$ photons and is encoded in the $X$ or $Y$ basis. Compared to the DQPS protocol, phase randomization allows treating each block as a mixture of Fock states and therefore enables a stronger security analysis \cite{ShunKawakami.2016}. The protocol also increases security as compared to BB84 because the eavesdropper cannot predict when a detection event is meant to occur within a block. Figures \ref{fig:applications:discrete:optics}c~and~d present the implementation of the phase-randomized DQPS QKD protocol and the comparison of the secure key rate with the BB84 protocol obtained with the same transmitter \cite{GeorgeL.Roberts.2017}.

\subsection{Multi-Protocol, Multi-Clock Rate QKD}
\begin{figure}[h!]
	\centering
	\includegraphics[width=\columnwidth]{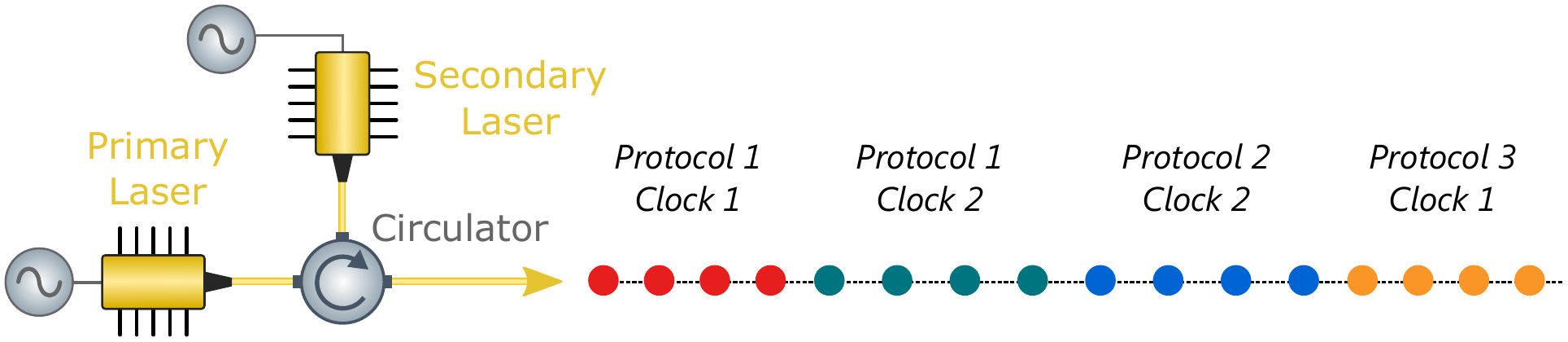}
	\caption{Concept of a multi-clock rate, multi-protocol QKD with a phase-seeded transmitter. The same transmitter can be used to prepare blocks of pulses encoded according to different protocols and at different clock rates, by only adapting the driving signals and without reconfiguration of the optical setup. With appropriate routing, a same QKD transmitter can communicate with different QKD receivers. }
	\label{fig:applications:multiprotocol}
\end{figure}

The above discussion shows that the same source is compatible with multiple protocols. While the operation clock-rate was not specifically discussed, the same source is in fact capable of encoding photon pulses at different clock rates with high versatility, within the bandwidth of the laser diodes. It is important to note that no modification of the setup is necessary to change clock or protocol (see Fig. \ref{fig:applications:multiprotocol}). This capability is promising for the deployment of QKD at large scale, where interoperability between hardware originating from different vendors will be essential.

\begin{figure}[h!]
	\centering
	\includegraphics[width=\columnwidth]{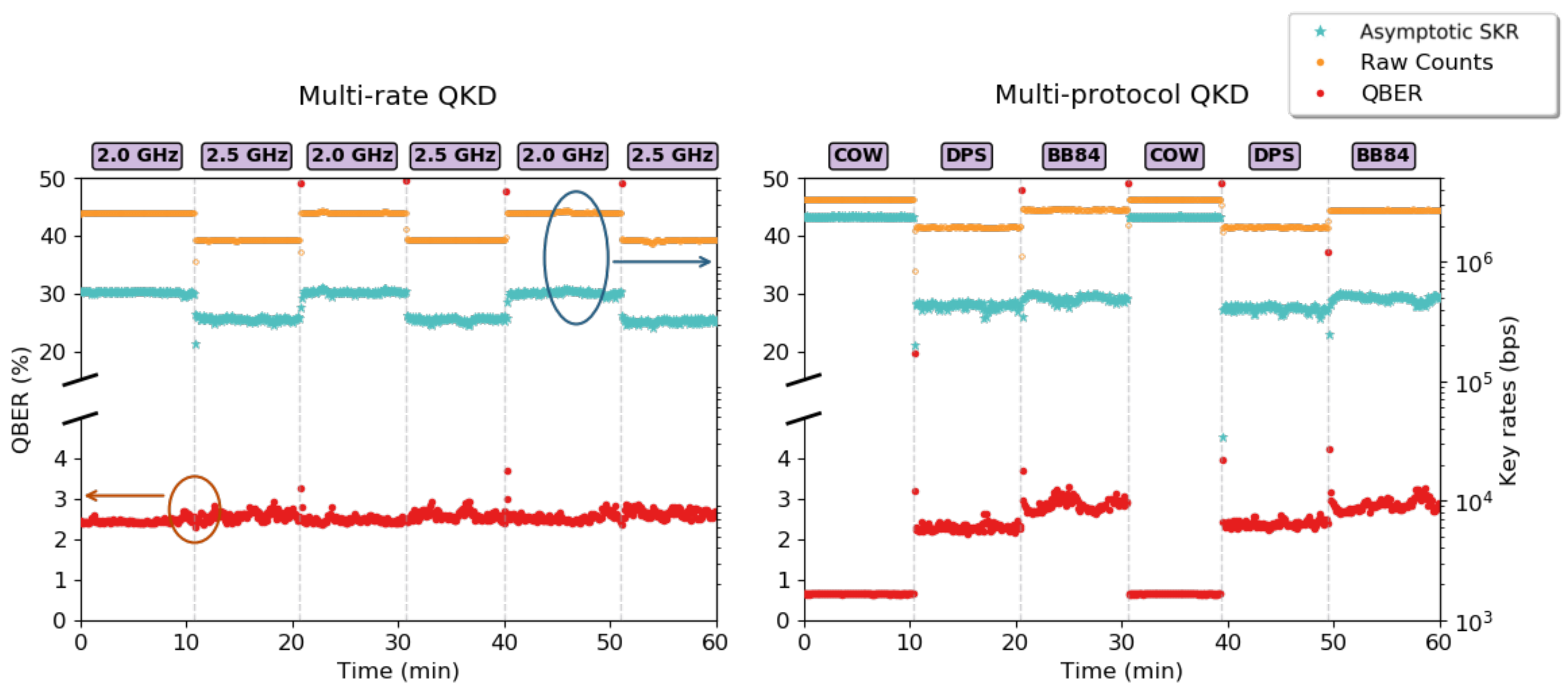}
	\caption{Experimental real-time multi-clock rate and multi-protocol QKD operation with a phase-seeded transmitter. Multi-rate QKD is demonstrated with the BB84 protocol, at 2 GHz and 2.5 GHz. Multi-protocol is demonstrated with the COW, DPS and BB84 protocols at 2 GHz clock rate. (\emph{Reproduced from Ref.~ \cite{Marco.2021}}).}
	\label{fig:applications:multiprotocol:expt}
\end{figure}

The phase-seeded QKD transmitter was shown to support real-time multi-protocol and multi-clock rate operation with 3 protocols (COW, DPS and BB84) and 2 different clock rates (2 GHz and 2.5 GHz) running sequentially without interruption (see Fig. \ref{fig:applications:multiprotocol:expt}).  The QKD transmitter proved able to adapt to protocol or clock-rate changes within a few seconds, which was only limited by the change of configuration of the driving electronic instruments \cite{Marco.2021}.

\subsection{Measurement-Device-Independent QKD}

An additional QKD protocol of great interest is Measurement-Device-Independent QKD (MDI-QKD)~\cite{Lo2012}. 
MDI-QKD eliminates all side-channels in the measurement devices, i.e. single-photon detectors, which are typically the most vulnerable components in a quantum communication system to side-channel attacks.
This is achieved through a novel system design where the two communicating parties, Alice and Bob, both encode and transmit random bits to a central relay node (``Charlie'') that performs a Bell state measurement and publicly announces the correlations. 
The users can thus post-select entangled states from these results to form a secure key, guaranteeing there has been no eavesdropper.

MDI-QKD effectively eliminates side-channels by equipping both users with laser transmitters and positioning the detectors in a node which can be completely untrusted.
This does, however, place more stringent requirements on the laser sources.
In particular, successful Bell state measurements require high-visibility two-photon (HOM) interference between pulses from Alice and Bob, demanding the pulses are highly indistinguishable in all degrees of freedom.
This is a major challenge when the laser transmitters are geographically remotely separated and feedback servo links for active stabilization are undesirable.
The problem is particularly acute at high clock rates, where information is encoded in short (10s ps) pulses and chirp / temporal jitter between pulses can strongly reduce indistinguishability~\cite{Comandar2016b}. 

\begin{figure}[h!]
	\centering
	\includegraphics[width=\columnwidth]{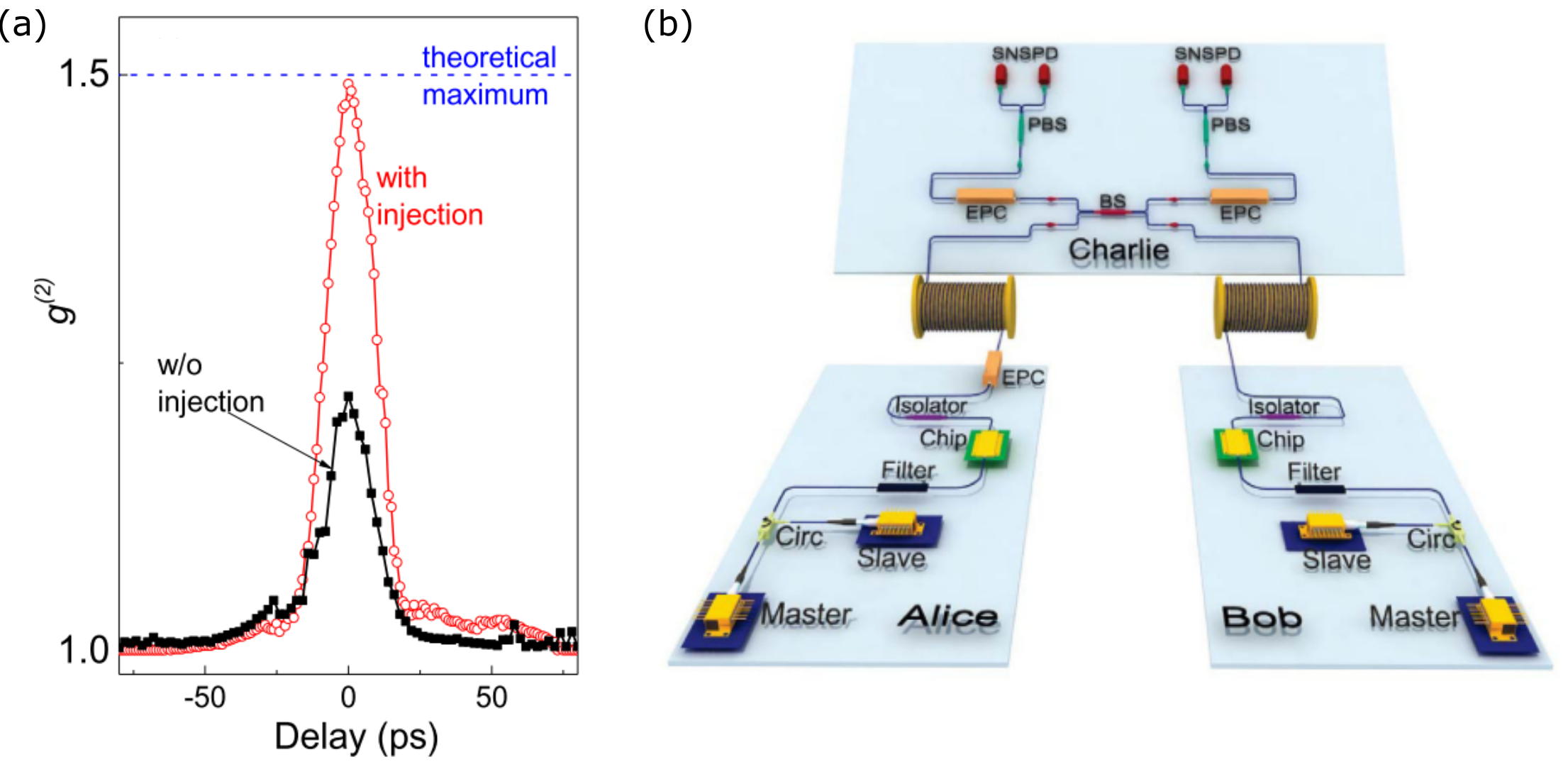}
	\caption{Application of directly modulated lasers to MDI-QKD: (a) second-order intensity correlation, $g^{(2)}$, measurement of interference between gain-switched lasers with and without optical injection (reprinted with permission from Ref.~\cite{Comandar2016b} \copyright The Optical Society); (b) schematic of complete MDI-QKD system where polarization-encoded bits are encoded by using an injection-locked gain-switched laser followed by a polarization modulator chip. (\emph{Reproduced from Ref.~\cite{Wei2019}}).}
	\label{fig:mdi}
\end{figure}

Fortunately, it has been shown that gain-switched OIL pulse generation techniques can solve this problem, enabling the use of laser sources for MDI-QKD at clock rates in excess of 1~GHz.
As described in Section~\ref{sec:theory}, OIL reduces the temporal jitter and relative intensity noise of gain-switched pulse trains, enabling precise temporal overlap when pulses from two such sources are interfered.
This is quantified by the two-photon visibility, also measured as the second-order intensity correlation function $g^{(2)}$.
Fig.~\ref{fig:mdi}(a) presents results from Ref.\cite{Comandar2016b}, demonstrating that interference between GHz gain-switched pulse sources typically achieves only $g^{(2)}\sim$1.2 (far from the theoretical optimum value of 1.5 for weak coherent states).
Such poor two-photon visibility would lead to unacceptably high error rates in MDI-QKD.
By exploiting OIL, however, the pulse chirp and jitter were significantly reduced (to a third of the original value~\cite{Comandar2016b}), enabling $g^{(2)}$ measurements approaching the theoretical optimum value of 1.5 for coherent states (corresponding to 50\% HOM visibility).

Both Comandar \textit{et al.}~\cite{Comandar2016} and Wei \textit{et al.}~\cite{Wei2019} have recently exploited this technique to demonstrate a complete MDI-QKD proof-of-principle system (Fig.~\ref{fig:mdi}(b)), with an impressively high 1~GHz and 1.25~GHz clock rate, respectively.
In both cases polarization encoding was used, with a polarization modulator included after the lasers to encode information onto the weak coherent pulses from the GS/OIL laser arrangement.
As a result of the enabled high clock rates and excellent HOM visibility, record MDI-QKD secure key rates were reported, including 1 Mb/s in the finite-size regime for metro-network scale distances~\cite{Comandar2016} and up to 30 b/s over 36~dB channel loss (180-km standard fiber)~\cite{Wei2019}.

\begin{figure}[h!]
	\centering
	\includegraphics[width=\columnwidth]{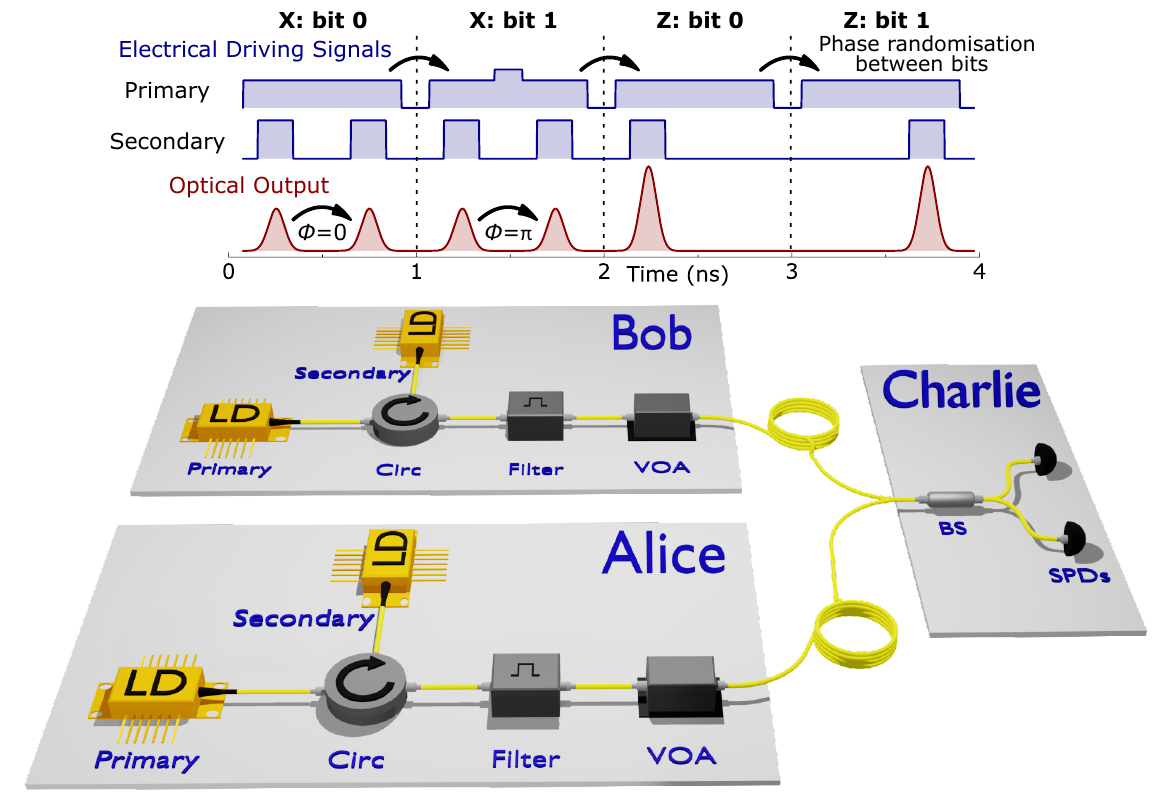}
	\caption{MDI-QKD system setup using gain-switched phase-seeded lasers to generate phase-encoded states directly from the lasers. This was achieved by modulating the primary / secondary laser waveforms, as shown in the top inset. (\emph{Reproduced from Ref.~\cite{Woodward2021a}}).}
	\label{fig:mdi2}
\end{figure}

A more recent advance has extended the concept of directly modulated lasers for MDI-QKD even further.
By employing time-bin encoding and applying modulator drive waveforms to both Alice and Bob's primary laser, i.e. similar to the transmitter for directly phase modulated BB84 described earlier, Ref.\cite{Woodward2021a} showed that direct laser modulation with GS/OIL could be used to generate all the required encoded states for MDI-QKD (Fig.~\ref{fig:mdi2}).
Thus, a complete MDI-QKD system was demonstrated without the need for polarization or phase modulators---significantly simplifying the setup.
Despite the random encoding of states applied to the lasers, they still resulted in 47\% HOM visibility, and operated at 1 GHz clock rate to achieve secure bit rates up to an order of magnitude greater than the previous state of the art.

\subsection{Twin-Field QKD}

A recent extension to the MDI-QKD concept, known as Twin-Field QKD (TF-QKD)~\cite{Lucamarini2018} offers the potential to enable quantum communication over even greater distances, surpassing even the repeaterless secret key capacity~\cite{Pirandola2017}.
Similar to MDI-QKD, the two communicating users in TF-QKD each encode and transmit coherent states to a central untrusted node where they are interfered and measured. 
A key difference, however, is that in TF-QKD information is encoded in the absolute phase of pulses generated by Alice and Bob, rather than the phase difference between two consecutive pulses from each user.
Measurements thus result in first-order interference, not two-photon interference, resulting in a quadratic improvement in key rate as a function of distance.

To implement TF-QKD, phase coherence is required between Alice and Bob's lasers---placing even more stringent requirements on their laser sources than for other QKD protocols. 
Phase fluctuations between users can result from both phase and optical frequency fluctuations of the user's lasers, degrading the interference visibility.
OIL has been shown as a solution to overcome this, however: by including a narrow-linewidth laser at Charlie with light transported to each user over a servo fiber.
Alice and Bob can then inject this seed light into their lasers to fix the wavelength and phase to that of the master laser, thus establishing a common shared phase reference~\cite{Fang2020}.
This technique has even been demonstrated for field trials of TF-QKD over 400~km~\cite{Liu2021}.
Additionally, it is noted that as with other QKD protocols, gain-switching has been shown to be an ideal technique in TF-QKD for generating pulses onto which information is encoded~\cite{Minder2019}.

\subsection{On-Chip Integration}
\begin{figure}[h!]
	\centering
	\includegraphics[width=\columnwidth]{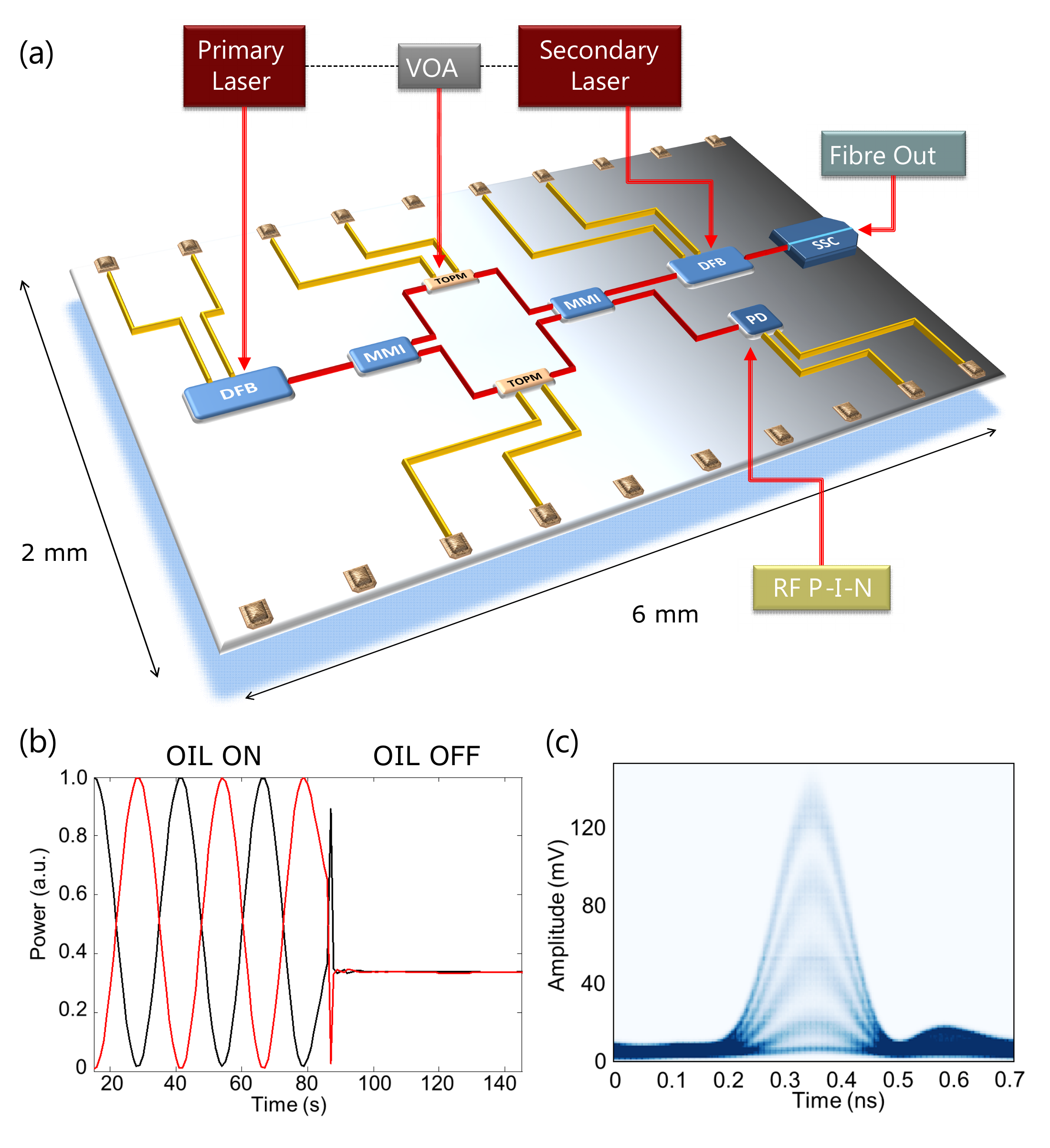}
	\caption{a Schematic diagram of the on-chip phase-seeded source for QKD applications. The circuit is manufactured on a 2~mm~$\times$~6~mm indium phosphide chip, allowing for both optically active and passive components to be integrated in the same substrate (VOA: variable optical attenuator; MMI: 2x2 multi-mode interferometer coupler; TOPS: thermo-optic phase shifter; PD: photodiode; SSC: spot-size converter). b Coherence transfer. Interference visibility of the secondary GS pulses measured at the output of the receiver's AMZI in the presence, absence of a CW seed from the primary laser. When OIL is ON, interferences are measured with 98.3 $\%$ visibility. When OIL is OFF the visibility is $<<1\%$, showing that there is no phase coherence between successive GS pulses. c Eye diagram of a RZ-8DPSK signal after demodulation showing 5 distinct intensity levels as expected. (\emph{Adapted from Ref. \cite{TaofiqK.Paraiso.2019})}.}
	\label{fig:applications:chip}
\end{figure}

The phase-seeded source is well suited for on-chip integration. In the perspective of mass deployment of QKD, photonic integration provides an attractive way to produce miniature and scalable optical hardware at low cost and with high reproducibility. Several QKD transmitter chips have been demonstrated in the recent years, showing that this will be a viable approach in the next development of the technology \cite{Sibson2017,Ma2016,Geng2019,Bunandar.2018,TaofiqK.Paraiso.2019,HenrySemenenko.2020,L.Cao.2020}. To fully benefit from this integration, the photonic chips should ideally be kept of low footprint, operate with low power consumption and the complexity of the driving electronics should be kept minimal. Integrating the phase-seeded source on chip reduces the use of electro-optic phase modulators, which are photonic components of large footprint and high power consumption. This modulator-free design was successfully implemented on a 2~mm~$\times$~6~mm chip, shown in Fig. \ref{fig:applications:chip}. Because of the absence of an optical isolator or circulator in conventional photonic integration platforms, an optical attenuator is used to suppress reciprocal seeding effects while still allowing efficient OIL from the primary laser into the secondary laser for precise control of the phase of the GS pulse train.

\begin{figure}[h!]
	\centering
	\includegraphics[width=\columnwidth]{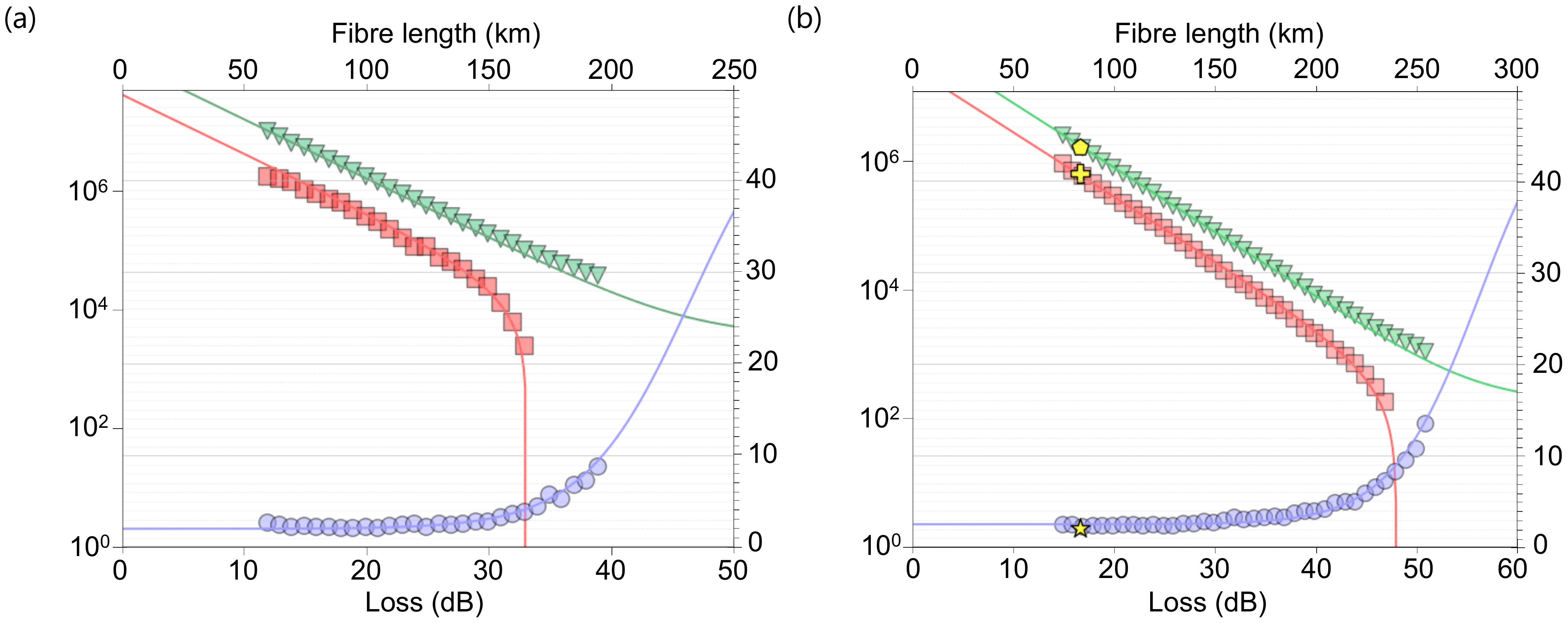}
	\caption{High-bit rate QKD. The performance of the phase-seeded transmitter chip is measured as a function of the channel loss emulated with a variable optical attenuator. Photons at the receiver are detected using SNSPDs with 90~$\%$ efficiency. QBER (blue circles), sifted key rate (green triangles) and secure key rate (red squares) are shown for (a) the DPS protocol and (b) the decoy-state BB84 protocol. Yellow symbols were acquired with a 75~km fibre spool. (\emph{Reproduced from Ref. \cite{TaofiqK.Paraiso.2019}}).}
	\label{fig:applications:chip:results}
\end{figure}

The phase-seeded transmitter chip was proven capable of high-bit rate QKD operation for both the decoy state BB84 and DPS protocols. Fig. \ref{fig:applications:chip:results} shows the performance of the chip for the DPS protocol and for the decoy-state BB84 protocol as a function of the channel loss. In this experiment, the CW laser linewidth was $<4.5$~MHz For the DPS protocol (fig. \ref{fig:applications:chip:results}~a), a QBER of 2.5~$\%$ and an asymptotic SKR of 400 kb/s were measured over a 20~dB-loss channel (equivalent to 100~km of standard SMF fiber). For the decoy-state BB84 protocol (fig. \ref{fig:applications:chip:results}~b), a QBER as low as 2.2~$\%$ and an asymptotic SKR of 270~kbps were measured over a 20~dB-loss channel~\cite{TaofiqK.Paraiso.2019}.

\section{Discussion \& Perspectives}
\label{sec:discussion}

The wide range of applications of QKD demonstrations employing directly modulated lasers, spanning all major protocols to date, is testament to the numerous benefits that the platform offers.
At the simplest level, using only gain switching of semiconductor lasers provides a compact source of pulses, capable of GHz clock rates.
Pulse performance can be significantly improved using OIL, with reduced jitter, chirp and relative intensity noise, in addition to enhanced modulation bandwidth.
For quantum communications, this can lead to reduced quantum bit error rates and thus, improved secure bit rates.
The enhancement is particularly marked for certain protocols such as MDI-QKD which rely on two-photon interference of pulses from communicating parties.
The opportunities of direct laser modulation go beyond simply generating pulses, however, with the demonstration of phase-seeding for the direct generation of amplitude and phase modulated pulses, offering time-bin encoded information straight out the laser system.
This is thus a significant simplification for quantum communications transmitter technology, paving the way to simpler, more compact systems and thus, more practical widespread deployments.

Looking to the future, it is also worth considering the scalability of laser modulation techniques for further advancing secure bit rate.
A major limitation for current QKD systems is the system clock rate, which has yet to exceed 5~GHz to date~\cite{Grunenfelder2020}, since higher clock rates approach the modulation bandwidth of commercial semiconductor lasers.
What are the opportunities then, for extending directly modulated laser scheme to higher clock rates?
Two primary challenges exist: (1) higher speed electronics which can generate electrical driving patterns with $<$100~ps modulation features are required; (2) semiconductor lasers with greater modulation bandwidth are needed.
Current commercial laser diodes are typically limited to $\sim$10~GHz gain-switching bandwidth, which is the primary challenge to be overcome, since increasing clock rates beyond a few GHz using current lasers will lead to inter-pulse correlations~\cite{Yoshino2018}, which violates the requirement for phase randomized bits.
There are reasons to be optimistic, however, with recent impressive demonstrations of high-bandwidth compact laser sources (e.g. $>$10s~GHz) through careful device engineering~\cite{Yamaoka2021} and further injection locking / feedback cavities~\cite{Liu2020}.

Finally, we note that the potential for exploiting directly amplitude and phase modulated lasers extends beyond QKD, with strong potential for broader impact in classical high-bandwidth communications, as well as other quantum technologies, such as quantum sensing and quantum computing.
Indeed, as quantum applications mature, there is a symbiotic relationship between quantum and laser technologies, with the former exploiting the decades of research and development in high-performance light source development, while simultaneously driving new laser research to meet the needs of emerging applications using quantum light.
This review has highlighted the promise of modulated semiconductor lasers, utilizing gain-switching, optical injection locking and direct phase modulation in particular, and we foresee a bright future for further exploitation and advancement of optical technologies in the quantum technology landscape.

\begin{acknowledgement}
This work has been funded by the Innovate UK project AQUASEC, as part of the UK National Quantum Technologies Programme. V.~L. acknowledges financial support from the EPSRC (EP/S513635/1) and Toshiba Europe Ltd.
\end{acknowledgement}

\providecommand{\WileyBibTextsc}{}
\let\textsc\WileyBibTextsc
\providecommand{\othercit}{}
\providecommand{\jr}[1]{#1}
\providecommand{\etal}{~et~al.}

\end{document}